\documentclass[usenatbib]{mnras}
\pdfoutput=1
\usepackage{mathptmx}
\usepackage{txfonts}

\usepackage[T1]{fontenc}
\usepackage{ae,aecompl}

\usepackage{graphicx}
\usepackage{longtable}
\usepackage[utf8]{inputenc}
\usepackage[english]{babel}

\newcommand{\Ms}{\ensuremath{M_{\odot}}}
\newcommand{\Zs}{\ensuremath{Z_{\odot}}}
\newcommand{\eg}{{\it e.g.}}
\newcommand{\cf}{{\it c.f.~}}
\newcommand{\ie}{{\it i.e.}}

\newcommand{\beq}{\begin{equation}}
\newcommand{\eeq}{\end{equation}}
\newcommand{\mtot}{\ensuremath{M_{\rm tot}}}
\newcommand{\mzams}{\ensuremath{M_{\rm ZAMS}}}
\newcommand{\mchirp}{\ensuremath{M_{\rm chirp}}}

\newcommand{\mbhone}{\ensuremath{M_{\rm BH1}}}
\newcommand{\mbhtwo}{\ensuremath{M_{\rm BH2}}}
\newcommand{\kmps}{\ensuremath{{\rm~km~s}^{-1}}}

\newcommand{\mcl}{\ensuremath{M_{cl}}}
\newcommand{\rh}{\ensuremath{r_h}}

\newcommand{\tmrg}{\ensuremath{t_{\rm mrg}}}
\newcommand{\taumrg}{\ensuremath{\tau_{\rm mrg}}}
\newcommand{\tej}{\ensuremath{t_{\rm ej}}}

\newcommand{\nmrgin}{\ensuremath{N_{\rm mrg,in}}}
\newcommand{\nmrgout}{\ensuremath{N_{\rm mrg,out}}}

\newcommand{\nnsbound}{\ensuremath{N_{\rm NS,bound}}}

\newcommand{\nbseven}{{\tt NBODY7~}}
\newcommand{\nbsix}{{\tt NBODY6 }}
\newcommand{\nbpp}{{\tt NBODY6++ }}
\newcommand{\bse}{{\tt BSE }}

\newcommand{\tevol}{\ensuremath{T_{\rm evol}}}
\newcommand{\fbin}{\ensuremath{f_{\rm bin}}}
\newcommand{\fobin}{\ensuremath{f_{\rm Obin}}}
\newcommand{\fgwp}{\ensuremath{f_{\rm GWp}}}
\title[Stellar-mass black holes in open clusters II]
{Stellar-mass black holes in young massive and open stellar
clusters and their role in gravitational-wave generation II}

\author[S. Banerjee]{
Sambaran Banerjee$^{1,2}$\thanks{E-mail: sambaran@astro.uni-bonn.de (SB)}
\\
$^{1}$Argelander-Institut f\"ur Astronomie (AIfA),
Auf dem H\"ugel 71, D-53121, Bonn, Germany\\
$^{2}$Helmholtz-Instituts f\"ur Strahlen- und Kernphysik (HISKP),
Nussallee 14-16, D-53115 Bonn, Germany
}

\pubyear{2017}

\begin{document}
\label{firstpage}
\pagerange{\pageref{firstpage}--\pageref{lastpage}} 
\maketitle

\begin{abstract}
The study of stellar-remnant black holes (BH) in dense stellar clusters is now
in the spotlight, especially due to their intrinsic ability to form
binary black holes (BBH) through dynamical encounters, that potentially coalesce
via gravitational-wave (GW) radiation. In this work,
which is a continuation from a recent study (Paper I),
additional models of compact stellar clusters with initial masses $\lesssim10^5\Ms$ and also
those with small fractions of primordial binaries ($\lesssim10$\%)
are evolved for long term, applying the direct N-body approach, assuming
state-of-the-art stellar-wind and remnant-formation prescriptions.
That way, a substantially broader range of computed models than that
in Paper I is achieved. As in Paper I, the general-relativistic BBH mergers
continue to be mostly mediated by triples that are bound to the clusters rather  
than happen among the ejected BBHs. In fact, the number of such in situ BBH mergers,
per cluster, tend to increase significantly with the introduction of a small
population of primordial binaries.
Despite the presence of massive primordial binaries, the merging BBHs,
especially the in situ ones, are found to be exclusively dynamically assembled
and hence would be spin-orbit misaligned.
The BBHs typically traverse through both the LISA's and the LIGO's detection bands,
being audible to both instruments. The ``dynamical heating'' of the BHs
keeps the Electron-Capture-Supernova (ECS) neutron stars (NS)
from effectively mass segregating and participating in exchange
interactions; the dynamically-active BHs would also exchange into
any NS binary within $\lesssim1$ Gyr. Such young massive and open clusters have the potential to
contribute to the dynamical BBH merger detection rate to a similar extent as their more massive
globular-cluster counterparts.
\end{abstract}


\section{Introduction}\label{intro}

The study of dynamical interactions of black holes (hereafter BH)
in dense stellar systems is now
nearly 30 years old and is still gaining momentum.
A key point of interest in such a process is 
its inherent potential of generating gravitational waves (hereafter GW) from
the general-relativistic (hereafter GR) inspiral of
dynamically-formed binary black holes (hereafter BBH;
\citealt{1993Natur.364..421K,1993Natur.364..423S,2000ApJ...528L..17P}).
The interest in this line of study
has naturally got rejuvenated after the landmark detections of GW, from
BBH inspiral events, by the Advanced Laser Interferometer Gravitational-Wave Observatory
(hereafter LIGO; \citealt{2016PhRvL.116f1102A}) detector, namely,
GW150914 \citep{2016PhRvL.116f1102A,2016ApJ...818L..22A}, GW151226 \citep{2016PhRvL.116x1103A},
and the ``trigger'' event LVT151012 \citep{2016PhRvL.116f1102A}. Especially,
the most recent LIGO detection, namely, the BBH inspiral event GW170104 \citep{Abbott_GW170104} 
prefer a spin-orbit
misalignment among the merging BHs (which may, moreover, be an anti-alignment;
see also \citealt{PhysRevX.6.041014,PhysRevX.6.041015,Farr_2017}),
provided the BHs' spins are large; the detection is, however, as well
consistent with zero or small BH spins.
A spin-orbit misaligned merger opens up the possibility that the members of
the merging BH pair might have been born independently
and paired up later, which is    
exactly what the dynamical scenarios achieve in an inherent way. 
Of course, spin-orbit misaligned BBH coalescences can, as well, be
potentially obtained via evolution of isolated, massive-stellar
binaries (see below).

In a nutshell, the dynamical formation of (stellar-mass) BBHs
in star clusters is a rather straightforward process:
if a certain number of BHs receive sufficiently low
natal kicks, during their formation through core collapse of massive
stars (\eg, in the case of a direct collapse or ``failed supernova''),
that they remain bound to the gravitational potential of the cluster,
they would segregate to the innermost regions of the cluster. Depending on
the number of BHs retained in the cluster at birth and their masses
(which is $\sim10$ to $\sim100$ times the cluster's average stellar mass depending
on their progenitor stars' wind; see below), the collection
of bound BHs might undergo a runaway mass segregation (mass-stratification
or Spitzer instability; \citealt{1987degc.book.....S}) to form
a central and highly dense subsystem of BHs, where they would continuously
interact. Otherwise, the dynamical friction due to the dense stellar
environment of the cluster would as well act to keep the BHs centrally concentrated
(as these BHs are much more massive than the normal stars, they
would always segregate towards the cluster's center simply via dynamical
friction, rather than being driven by the two-body relaxation among the
average stars). In either case, it should be borne in mind that
as long as the BHs retain in the cluster, they continue to
interact with the rest of the stellar entities and not only among themselves, \ie,
they do \emph{not} comprise a so-called ``dynamically isolated subsystem''
\citep{Morscher_2013,Breen_2013}.
Such interactions happen, both, in the form of close encounters between
the BHs (or the BBHs) and the individual stars or stellar binaries (\eg, tidal capture,
exchange) and of the overall interaction between the BHs with the stellar
background (the dynamical friction).

Such a dense BH core serves as a factory for dynamically
pairing BBHs, often via the three-body mechanism \citep{1987degc.book.....S,2003gmbp.book.....H}.
If primordial stellar binaries are present, they would as well mediate
the pairing process (see, \eg, \citealt{Rodriguez_2016,2016arXiv160300884C}).
The subsequent frequent and super-elastic \citep{1987degc.book.....S}
encounters of a BBH with other single
BHs and BBHs serve as a recipe for (a) injecting kinetic energy (K.E.)
into the BH sub-cluster and as well into the whole star cluster causing the
latter to expand \citep{2007MNRAS.379L..40M,2008MNRAS.386...65M},
(b) ejecting single and binary BHs from the cluster depleting the
BH population, and (c) forming triple-BH systems within the BH-core. The ejected
BBHs are typically dynamically tightened (hardened; \citealt{1975MNRAS.173..729H})
and also eccentric, an adequate combination of which \citep{Peters_1964} would lead to the inspiral
via GW radiation and the coalescence of a BBH within the Hubble time.

The triple-BHs that are bound to the clusters, on the other hand,
would typically undergo a large eccentricity boost and hence potentially GW inspiral and coalescence,
of the inner binary. If the triple is hierarchical and sufficiently long lasting, this would be due to  
the classical/secular eccentric Kozai oscillations
\citep{Kozai_1962,Miller_2002,Wen_2003,Lithwick_2011,Antonini_2014,Hoang_2017}.
However, triples, that trigger eccentricity boost and hence GR coalescence,
can as well be intermediate, metastable (or democratic) systems which
disintegrate (if GR effects are excluded), typically, after a few internal orbits
of the triple; see, \eg, \citet{Hut_1983,Samsing_2014}. The latter type of triples can be either 
an intermediate phase of a resonant BBH-BH interaction or an outcome of a close
interaction between two BBHs of similar binding energies. Hierarchical, stable
configurations, on the other hand, are more likely to arise during interactions
between two BBHs where one is much tighter than the other.
A sufficient eccentricity boost of the triple's constituent BBH, either
during a metastable, resonant interaction phase \citep{Samsing_2014}
or due to a regular Kozai oscillation,
would initiate its GR inspiral, provided this happens before the triple gets either disintegrated
through its internal dynamical evolution or perturbed by an intruder. 
The eccentricity boost is instrumental here; the GR
inspiral and merger of a sufficiently tight, (nearly) circular BBH is practically
impossible while being bound to a stellar cluster since a BBH
would get ejected due to dynamical interactions well before it achieves
the required tightness (but this may be possible in galactic nuclei where the escape
speed is typically $>100\kmps$). In other words, although a star cluster
continues to eject single and binary BHs and form BH triples until its BH reservoir is (nearly) depleted,
the occurrence of a dynamical BBH inspiral is quite a coincidence
but is inherent to the dense stellar environment.

Since 1990s, aspects of the above mechanism is studied at various
levels of detail. Following preliminary but pioneering studies such as
\citet{1993Natur.364..421K,1993Natur.364..423S,2000ApJ...528L..17P},
more recent direct N-body (\eg, \citealt{2010MNRAS.402..371B,2012MNRAS.422..841A,2013MNRAS.430L..30S,Ziosi_2014,Kimpson_2016,Wang_2016,2016PASA...33...36H,Mapelli_2016,Baumgardt_2017,Banerjee_2017,Park_2017})
and Monte-Carlo (\eg, \citealt{Downing_2010,Downing_2011,Morscher_2015,Rodriguez_2016,Rodriguez_2016a,2016arXiv160300884C,2016arXiv160906689C,Askar_2016})
calculations of model stellar clusters study the dynamically-driven depletion of
BHs, the resulting feedback onto the cluster and the dynamically-induced BBH inspirals
self consistently and in much more detail. Adopting somewhat simpler
but realistic conditions, detailed semi-analytic studies of these aspects
have also been performed recently \citep{Breen_2013,Breen_2013a,ArcaSedda_2016}.
Also, such a semi-analytic modelling, in the context of $\sim10^7\Ms$ nuclear stellar clusters,
has been recently conducted by \citet{Antonini_2016} and the possibility
of BBH coalescences resulting from triple stars in the field has been
studied by \citet{Silsbee_2017,Antonini_2017}.
By nature, Monte-Carlo calculations
are restricted to massive model clusters, typically of $\sim 10^5\Ms - 10^6\Ms$, that
are representatives (or progenitors) of classical globular clusters (hereafter GC). On the other
hand, due to high computational cost, direct N-body calculations are typically done with
clusters of $\lesssim10^5\Ms$. However, in the ``Dragon Simulations''
\citep{Wang_2016}, clusters that are relatively extended, with half-mass radii of $\approx3-8$ pc,
but comprising as many as $N\approx10^6$ stars (they can also be taken as representatives
of galactic nuclear clusters) are evolved, with the direct N-body program \nbpp \citep{Wang_2015},
for nearly a Hubble time on large GPU compute clusters.

The typical conclusion from such studies
is that the local dynamical BBH inspiral rate (density) is $\approx 5-10 {\rm~yr}^{-1} {\rm~Gpc}^{-3}$,
that would contribute several 10s of BBH inspriral detection per year
with the LIGO, given its proposed full sensitivity \citep{2010MNRAS.402..371B,Rodriguez_2015,
Askar_2016,Banerjee_2017,Park_2017}.
However, taking into account lower-mass, open-type clusters,
\citet{Banerjee_2017} anticipate a full-sensitivity dynamical BBH detection
rate up to $\sim 100$ per year.
In the majority of these numerical studies (but see \citealt{Banerjee_2017}; below),
most of the GW coalescences have occurred among BBHs
that are dynamically ejected from the clusters. Another important corollary is
that the BH population in massive clusters, although decaying monotonically with
time due to the dynamical interactions and the resulting ejections, it never actually gets completely depleted
even in a Hubble time, so that a substantial population of BHs would be retained
even in old GCs. In contrast, in the earliest studies (\eg, \citealt{1993Natur.364..421K}),
it was conceived that all except one or two BHs would be ejected out of an
old cluster, making them unpopular hosts of stellar-remnant BHs. Essentially,
such a long-term retention is due to the
energy generation through dynamical encounters in the BH-core (see above) that results
in substantial expansion of both the parent cluster and the BH-core itself
(see, \eg, Fig.~6 \& 7 of \citealt{Banerjee_2017}),
suppressing the BH-BH interaction rate. This ``self-regulation'' causes
the BH population to decline but exponentially;
such a self-regulatory behaviour is essentially a manifestation of the \citet{Henon_1975}
principle, according to which the central (dynamical) energy generation of a (post-core-collapse) cluster is controlled
by the energy demands of the bulk of the cluster. This principle is analogously
applicable to the energy generation due to dynamical encounters within the BH-core inside a stellar cluster,   
as done in the semi-analytic study by \citet{Breen_2013}.
This behaviour is consistent with the recent identification
of stellar-mass BH candidates in Galactic GCs (see below).

\begin{figure*}
\centering
\includegraphics[width=8.5cm,angle=0]{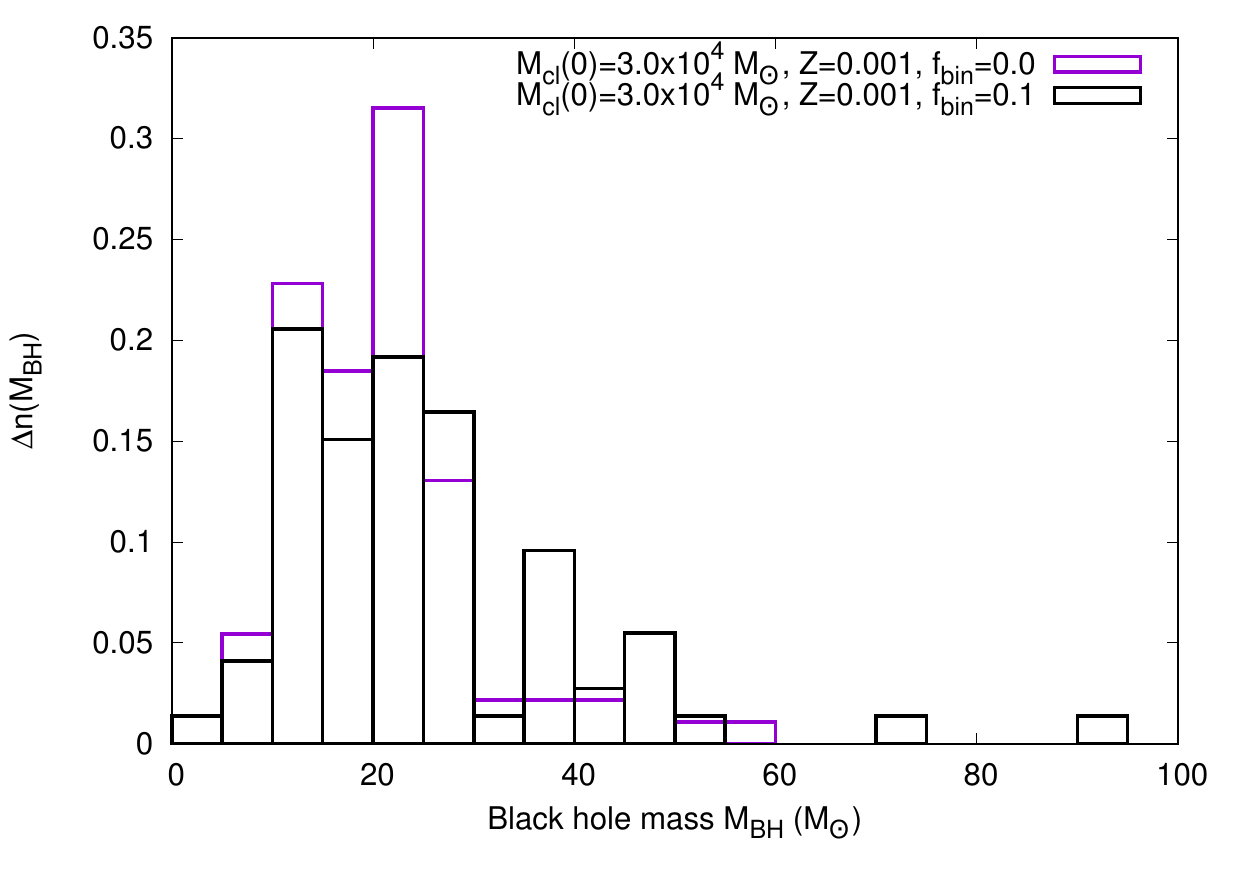}
\includegraphics[width=8.5cm,angle=0]{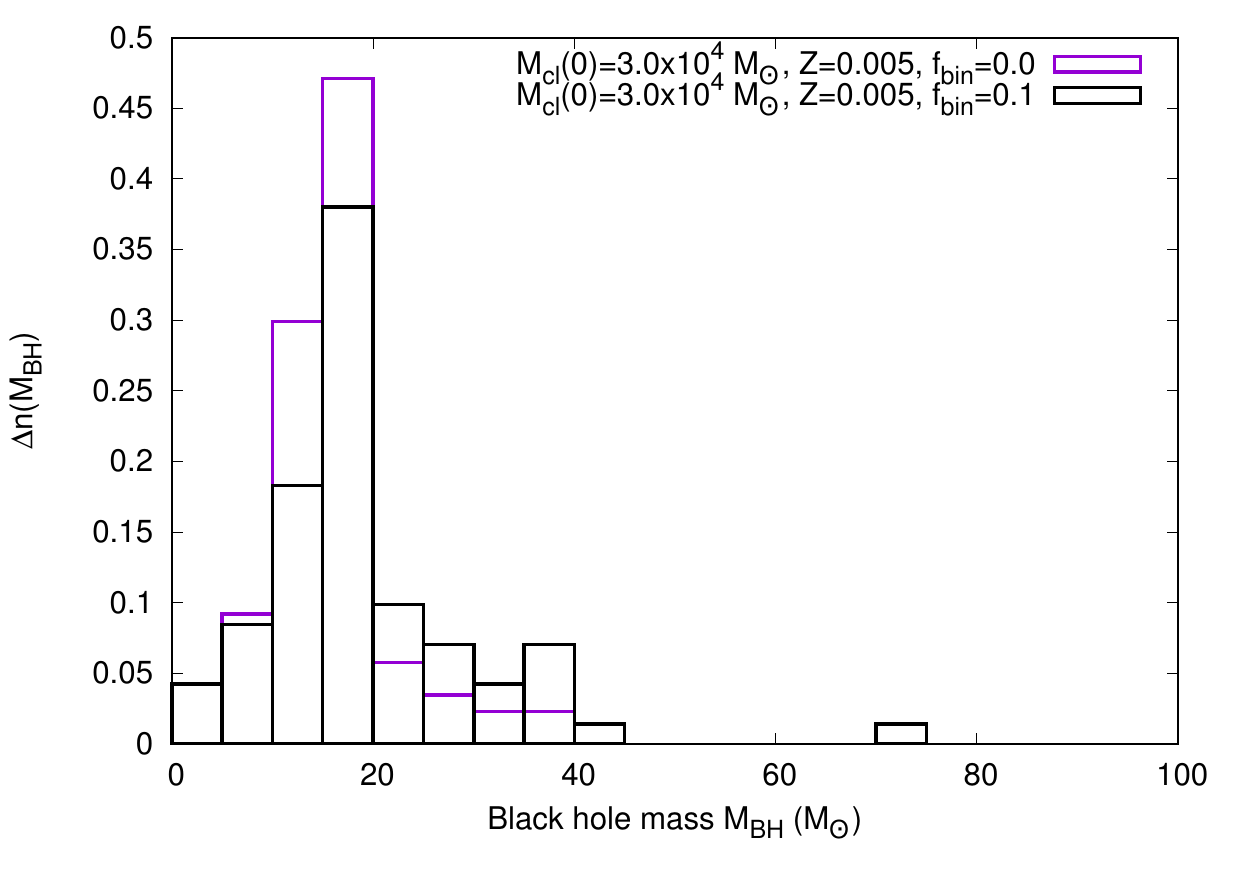}
	\caption{The (normalized) mass distributions of the BHs formed in the $\mcl(0)\approx3.0\times10^4\Ms$
	models with initially only single stars (blue-lined boxes) and with
	$\fbin\approx10$\% \emph{overall} primordial binaries (black-lined boxes), for
        $Z=0.05\Zs$ (left) and $0.25\Zs$ (right). The modified BH mass distributions
	in the primordial-binary cases are the combined outcomes of the internal
	evolution and dynamical modifications of the massive primordial binaries.
	Given that the BH-progenitor stars, in the models with primordial binaries,
	have nearly $\fobin\approx100$\% primordial binary fraction (as opposed to the much
	lower overall binary fraction; see Sec.~\ref{calc}), all the BHs in such
	models are either born in a binary or/and have their masses affected by
	(dynamically-modified) binary evolution. The typical resultant is a wider
	BH mass distribution, due to the massive binaries' mergers, CE phases, and
	tidally-enhanced stellar winds.
	}
\label{fig:mdist_BH}
\end{figure*}

\begin{figure*}
\centering
\includegraphics[width=8.5cm,angle=0]{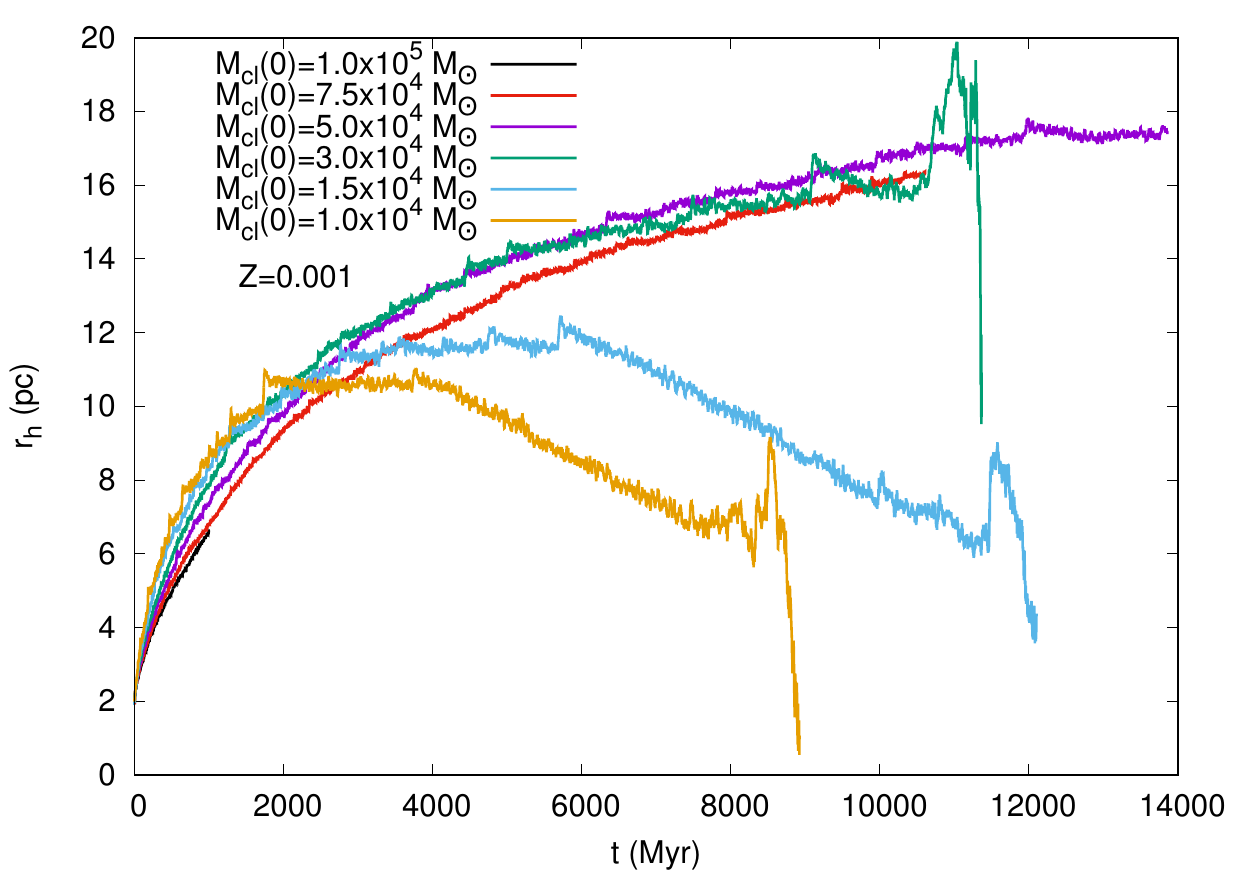}
\includegraphics[width=8.5cm,angle=0]{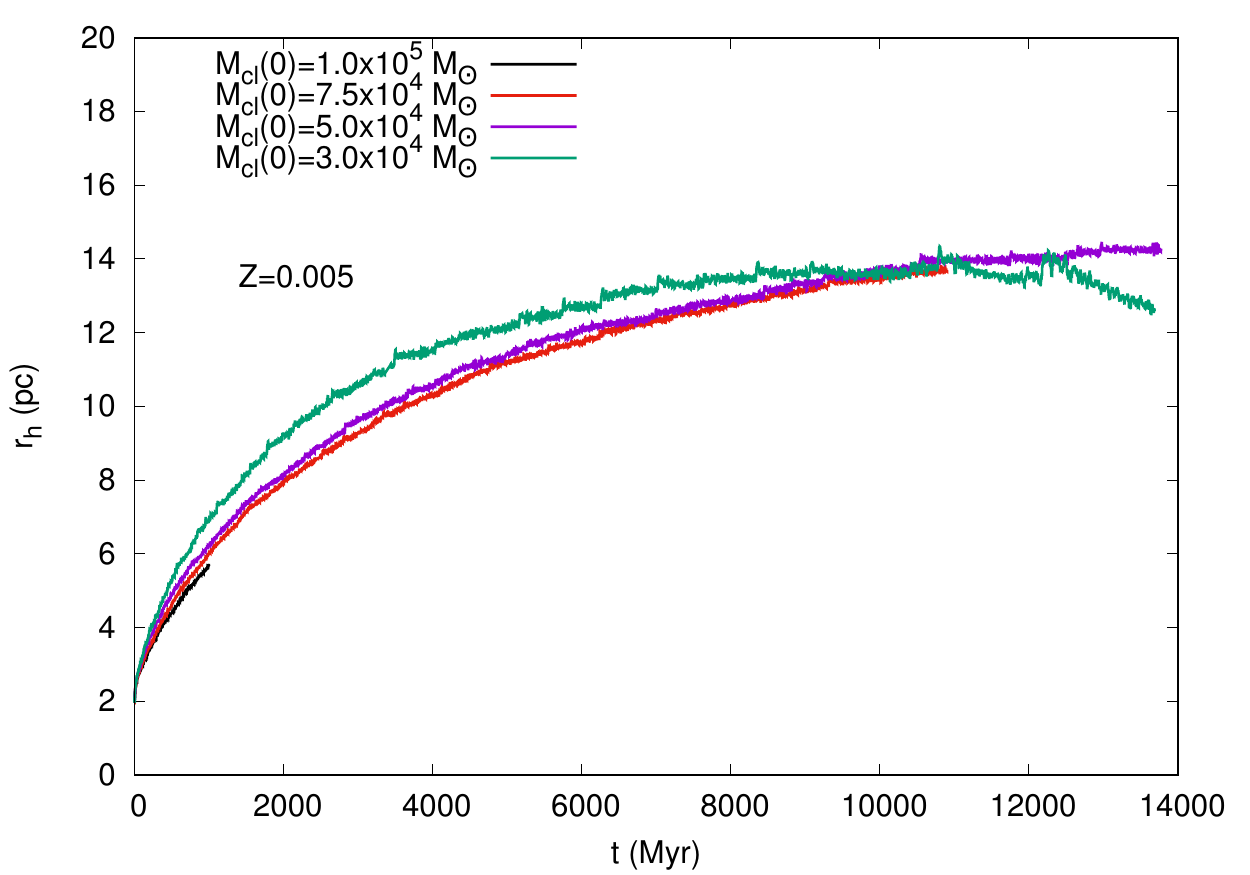}
\includegraphics[width=8.5cm,angle=0]{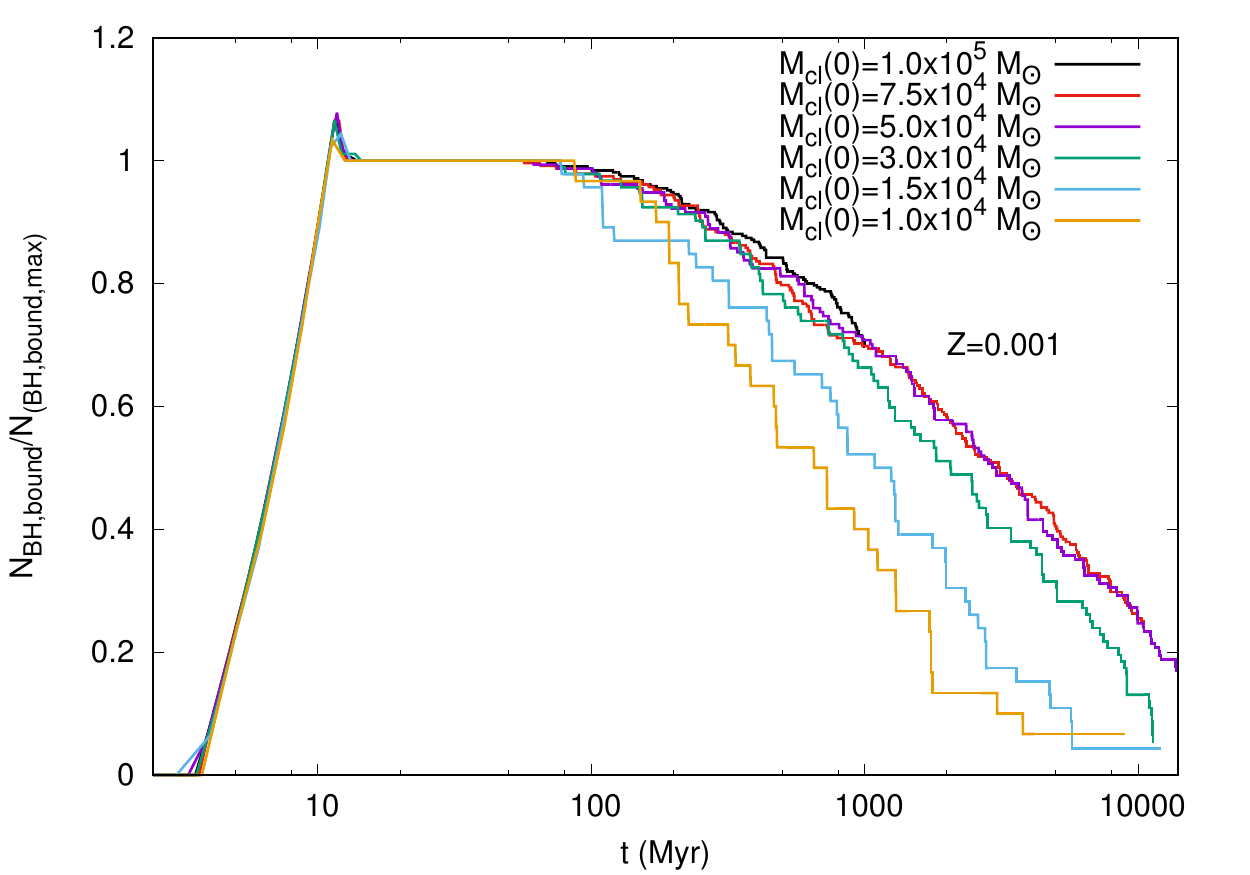}
\includegraphics[width=8.5cm,angle=0]{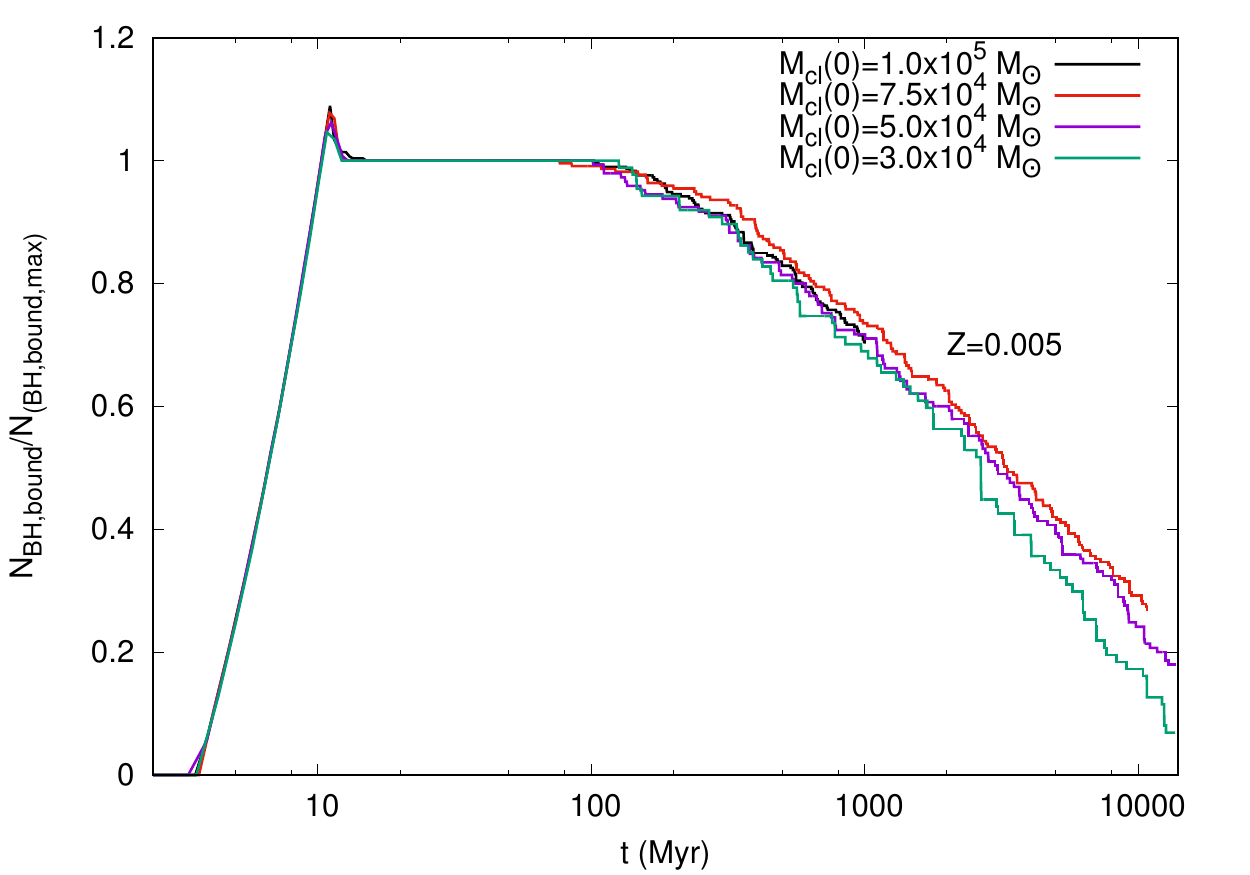}
\caption{{\bf Top panels}: The time evolution of the half-mass radius, $\rh$,
	for the computed initially single star-only
	models (see Table~\ref{tab1}) with metallicities $Z=0.05\Zs$ (left) and $0.25\Zs$
	(right). {\bf Bottom panels}: The time evolution of the fraction of
	BHs (normailzed w.r.t. the number of BHs that initially
	remain bound to the cluster, \ie, w.r.t. the flat part of the
	curves spanning between $\approx10 - \approx 100$ Myr) retaining in the
	models with these $Z$s.}
\label{fig:comp_all}
\end{figure*}

Space-based X-ray observation has been the most popular approach to identify
BH candidates in GCs, \eg, \citet{Maccarone_2007,Brassington_2010,Lutzge_2016}.
However, the best stellar-mass BH candidates have
been identified only recently by comparing X-ray and radio fluxes
from localized sources in the GCs M22 \citep{Strader_2012} and 47 Tuc \citep{Miller_Jones_2015}.
Given the rather
small probability of formation of BH X-ray binaries, the presence of such candidates
may indicate the existence of a much larger population of stellar-mass BHs
in GCs (see \citealt{Strader_2012} and references therein). In fact, the
recently-discovered very high mass-to-light-ratio
GCs or ``dark star clusters'' (DSC; \citealt{Banerjee_2011,Taylor_2015,Bovill_2016,Sollima_2016})
provide additional indications of the possible presence
of large populations of dark remnants in GCs.

Alternatively, BBHs can be formed through the evolution of isolated massive-stellar
binaries in the field. Such scenarios stem on either common-envelope (CE) 
evolution \citep{Belczynski_2016} or chemically-homogeneous evolution
\citep{DeMink_2009,DeMink_2016,Marchant_2016} of massive binaries.
In fact, GW170104 event is still consistent with no or small BH spins (\ie, with non-detection of BH spins),
in which case it can be reproduced through isolated massive-binary evolution also
(but preferably with modified physics; see \citealt{2017arXiv170607053B});
a spin-orbit misalignment (but preferably positive) can also be reproduced
from such a channel \citep{Belczynski_2016,2017arXiv170607053B,Oshaughnessy_2017a}. 
In such studies, BBH inspiral detection rate of $\sim10-\sim1000{\rm~yr}^{-1}$
has been estimated for the LIGO (at its proposed full sensitivity; \ie, up to 2 orders
of magnitudes higher detection rate than what is estimated for dynamically-formed BBHs).
Tight BBH formation through massive-binary
evolution in isolation, and hence their merger rate,
depends on the nature of CE evolution, tidal interactions among
massive stars, and their structural details,
all of which are still poorly understood. 
In that sense, the dynamical mechanism is, so far, the most assured
channel for BBH formation and their mergers.

Recently, \citet[hereafter Paper I]{Banerjee_2017} has computed
a preliminary set of evolutionary models of stellar clusters
with initial masses $1.0\times10^4\Ms-5.0\times10^4\Ms$,
half-mass radii 1-2 pc, that is typical for young massive clusters (hereafter YMC) and starburst clusters,
and having metallicities between $0.05\Zs-\Zs$.
Including the state-of-the-art stellar-wind and remnant-formation schemes of \citet{Belczynski_2010},
that allows the formation of direct-collapse (or failed-supernova) black holes (hereafter
DCBH; \citealt{Belczynski_2008,Spera_2015}),  
these models were evolved, all the way from their young age until their dissolution or at least 10 Gyr age,
for the first time with the direct N-body technique (utilizing \nbseven, a descendant of \nbsix; see Sec.~\ref{nbprog}),
where all sorts of dynamical encounters are treated completely and self-consistently. These comprise the most
advanced simulations of the dynamics of BHs in a star cluster to date. The motivation for the study
was that with decreasing metallicity, not only the average BH mass increases but also their
mass spectrum widens \citep{Belczynski_2010} --- an effect that is particularly
pronounced for the adopted stellar-wind
and remnant-formation recipes that comfortably include all the
LIGO-detected BHs \citep{2016ApJ...818L..22A}.
This would, in turn, modify the nature of the
dynamical interactions among the BHs and hence the BBH production.

An interesting fact, that became apparent from the computations in Paper I is that, in 
general and irrespective of the cluster's metallicity,
the number of (triple-induced; see above) BBH mergers happening while being
bound to the model cluster well exceeds that happening among the ejected BBHs
(see Table~1 of Paper I). This is in contrast to what
Monte-Carlo calculations of similarly and more massive systems but with similar model ingredients
(\eg, \citealt{Morscher_2015,Rodriguez_2016,2016arXiv160300884C,Askar_2016}),
have found and also to the outcomes of direct N-body calculations with equal- or binary-mass
BHs (\eg, \citealt{2010MNRAS.402..371B,Park_2017}). As discussed in 
the Sec.~3.2 of Paper I, this difference could be partly due to the
way in which compact subsystems are treated in Monte Carlo-based models
and partly due to the broad BH mass functions adopted in Paper I, that
has not been properly explored through the direct N-body method until Paper I
(but see \citealt{Park_2017}). 

The present work is a continuation of the study started in Paper I,
where additional model computations are 
performed, resulting in a broader mass spectrum of the model star clusters.
Furthermore, a preliminary set of models, that include a primordial-binary
population, is computed. Due to a broader BH mass spectrum, and hence the
potential for more interesting dynamics among the BHs, only lower metallicities,
of $Z\leq0.50\Zs$, are considered for the newer models. This naturally
places the majority of the models in dwarf galaxy-like hosts
(see also \citealt{Oshaughnessy_2017b} in this context); the most
massive ones can as well be looked upon as progenitors of present-day GCs
which, on an average, are of sub-$\Zs$ \citep{Harris_1996} or as parents of open clusters
which are also often of sub-$\Zs$ (\eg, \citealt{Heiter_2014}), in both Galactic and
extragalactic environments. In any case, the computations here suggest that
(Table~\ref{tab1}, Secs.~\ref{calc} \& \ref{res})
the number of BBH mergers, for a given cluster, does not
necessarily follow any systematics with the cluster's metallicity ($Z$) over a
wide range; in fact
higher-$Z$ clusters often produce as many or even more mergers than
their lower-$Z$ counterparts. Therefore, the general conclusions from this
work (Sec.~\ref{summary}) are likely to hold for clusters up to $\Zs$ and hence in all
types of hosts.
The lower-mass YMCs and open-type clusters, as considered here, are
as well interesting due to the fact that they comprise the vast majority
of the dense stellar clusters and would potentially serve as a competent
addendum to the GCs, in terms of their contributions to the dynamical
BBH mergers (Sec.~\ref{mrgrate}).

This paper is organized as follows.
The newer cluster models with primordial binaries are introduced in Sec.~\ref{calc},
along with a brief description of the \nbseven program in Sec.~\ref{nbprog}. The results from
the computations are elaborated in Sec.~\ref{res}: Sec.~\ref{bbh} discusses
the dynamically-formed BBH mergers' characteristics and their dependence on
the parent cluster's properties, Sec.~\ref{nsseg} discusses the behaviour
of the neutron stars in the presence of a dynamically-active population of BHs,
and Sec.~\ref{mrgrate} discusses the contribution of young massive and open clusters
to the dynamical BBH merger rate. Sec.~\ref{summary} recapitulates the results. 

\begin{figure*}
\centering
\includegraphics[width=8.5cm,angle=0]{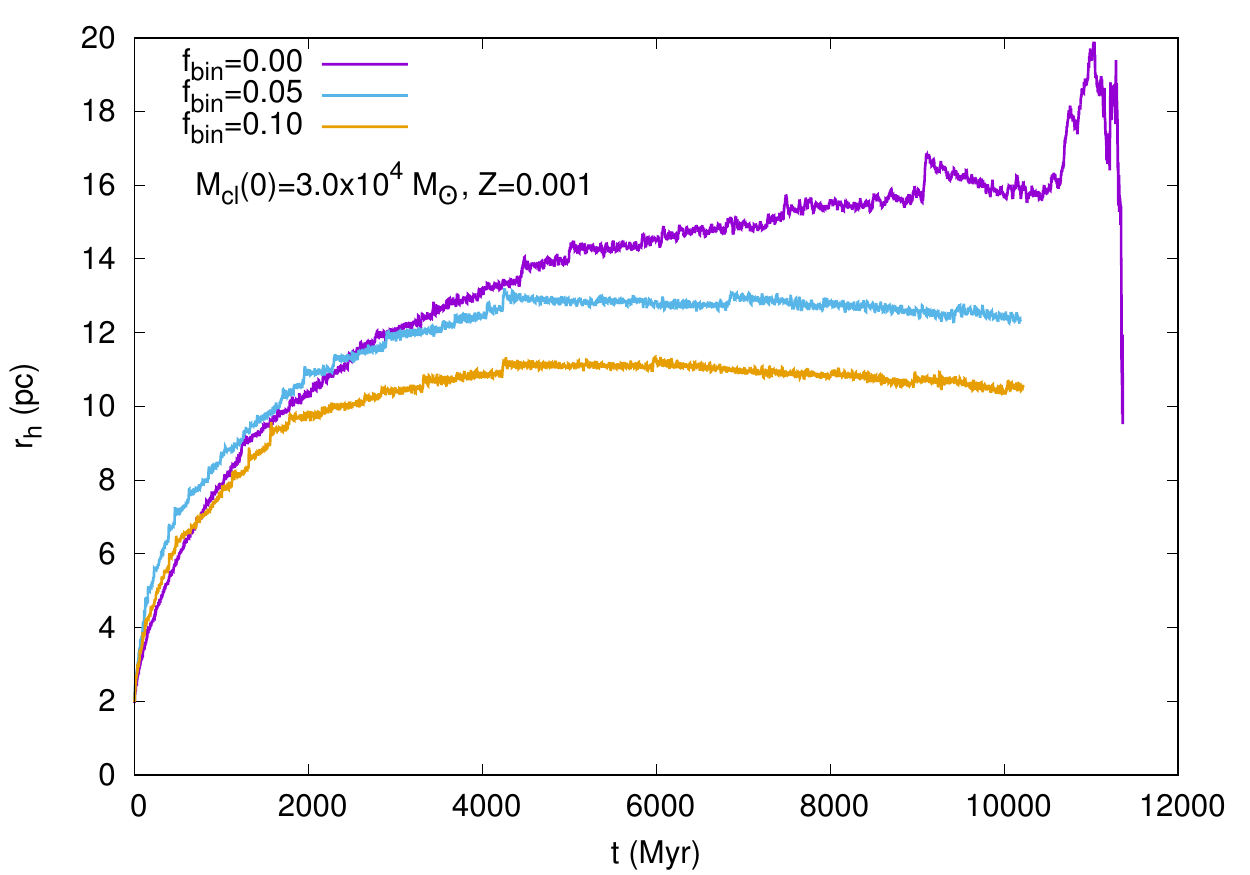}
\includegraphics[width=8.5cm,angle=0]{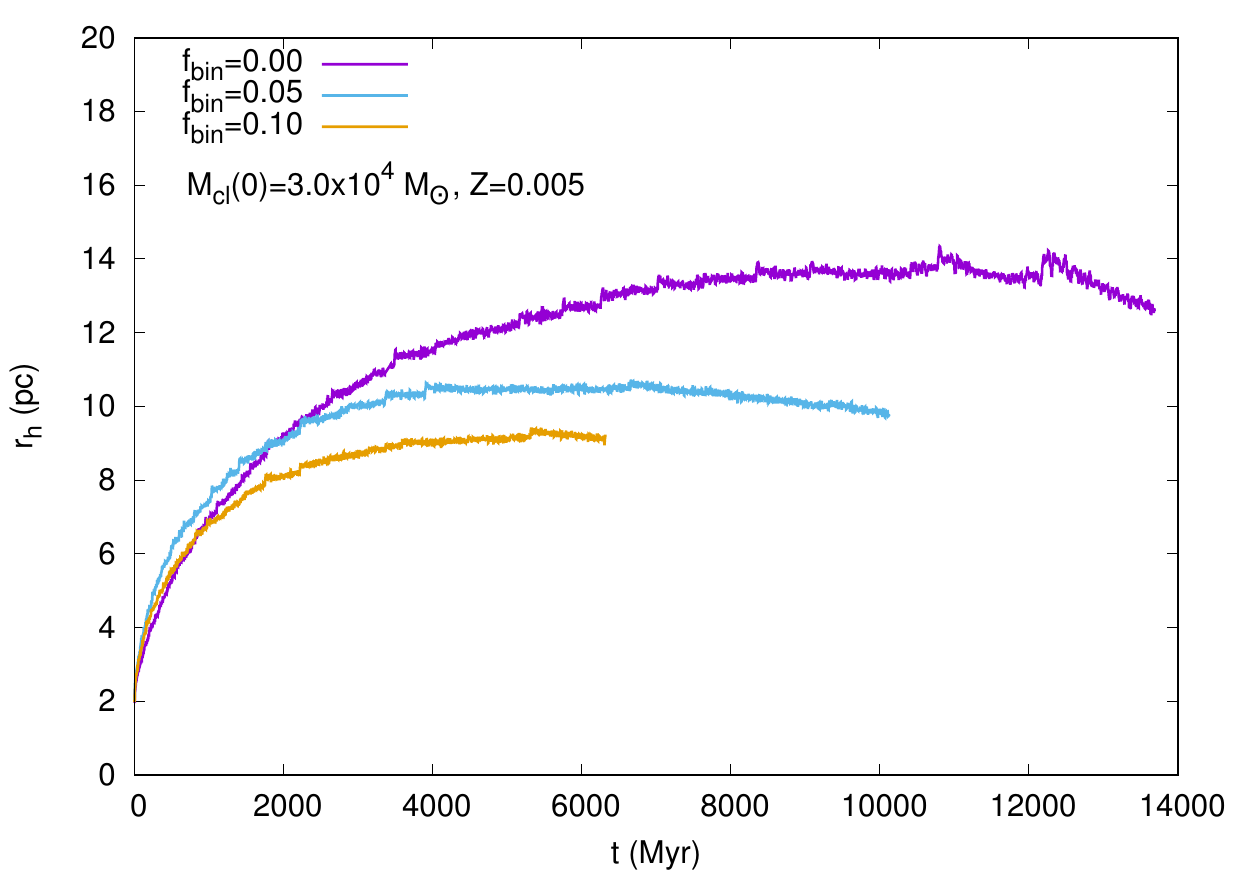}
\includegraphics[width=8.5cm,angle=0]{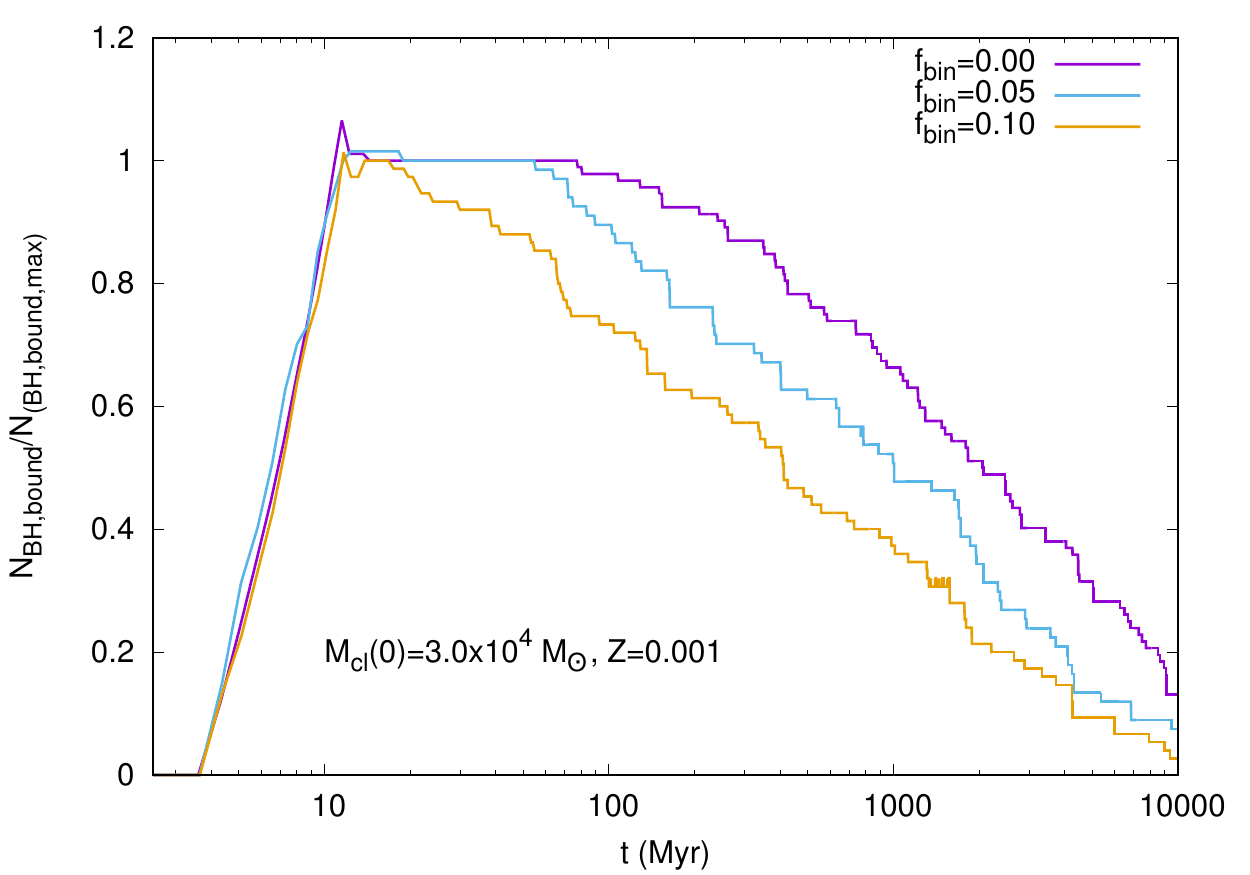}
\includegraphics[width=8.5cm,angle=0]{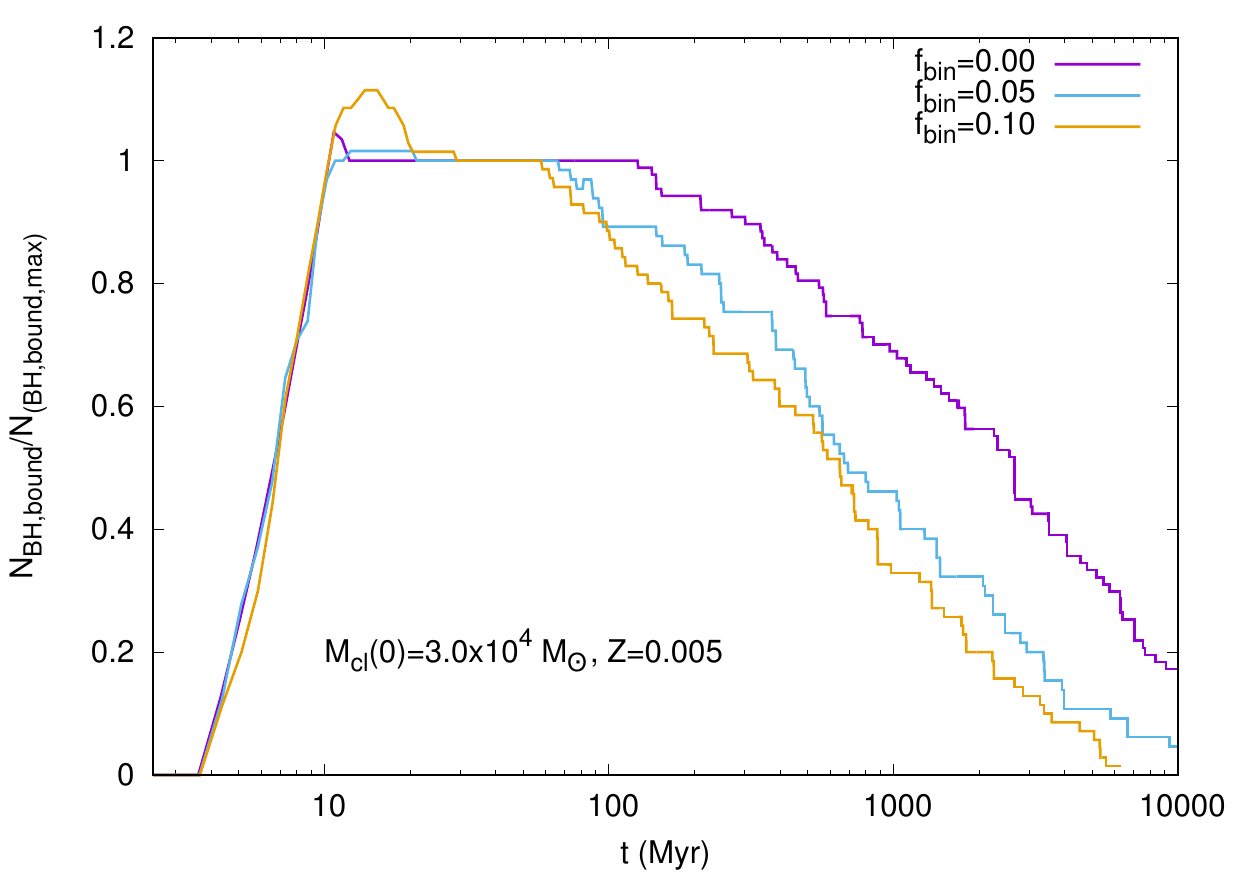}
\includegraphics[width=8.5cm,angle=0]{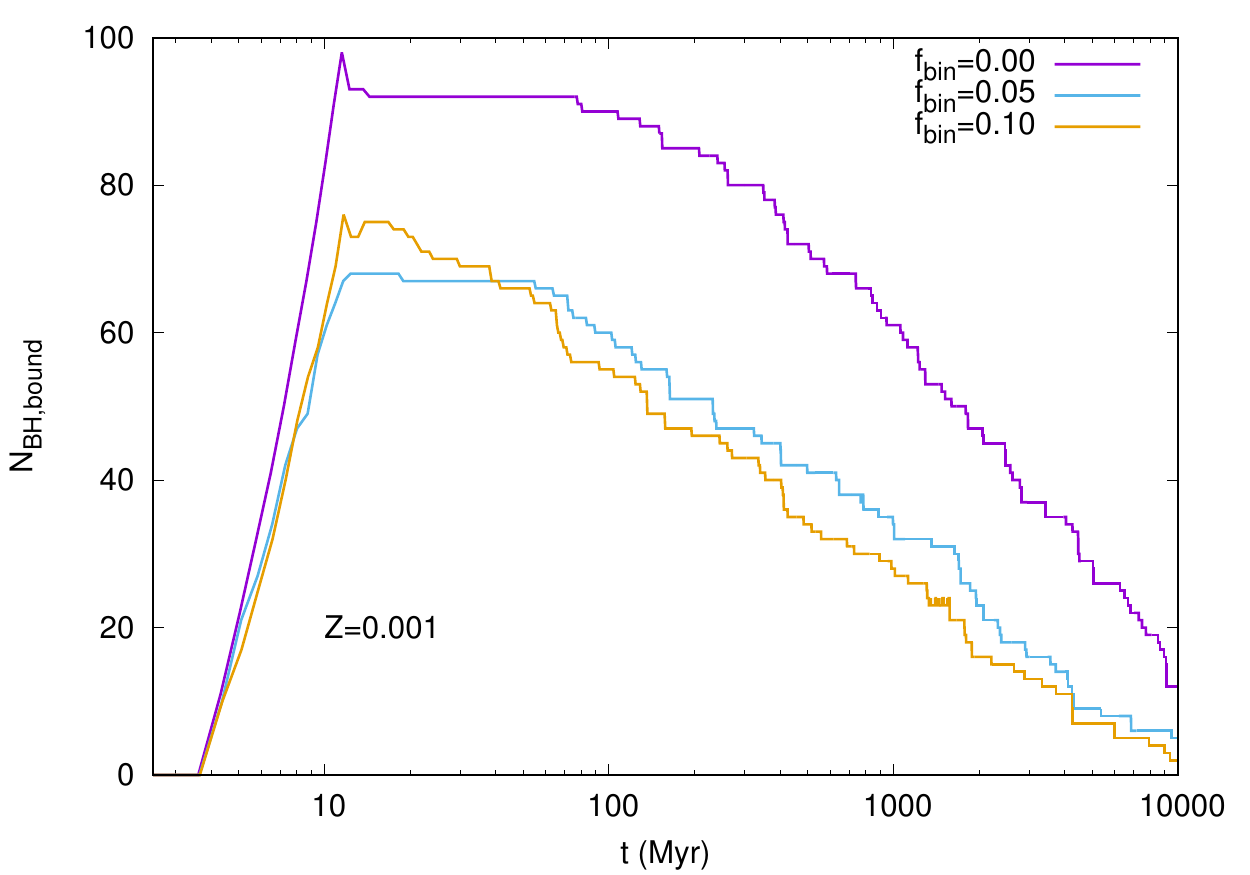}
\includegraphics[width=8.5cm,angle=0]{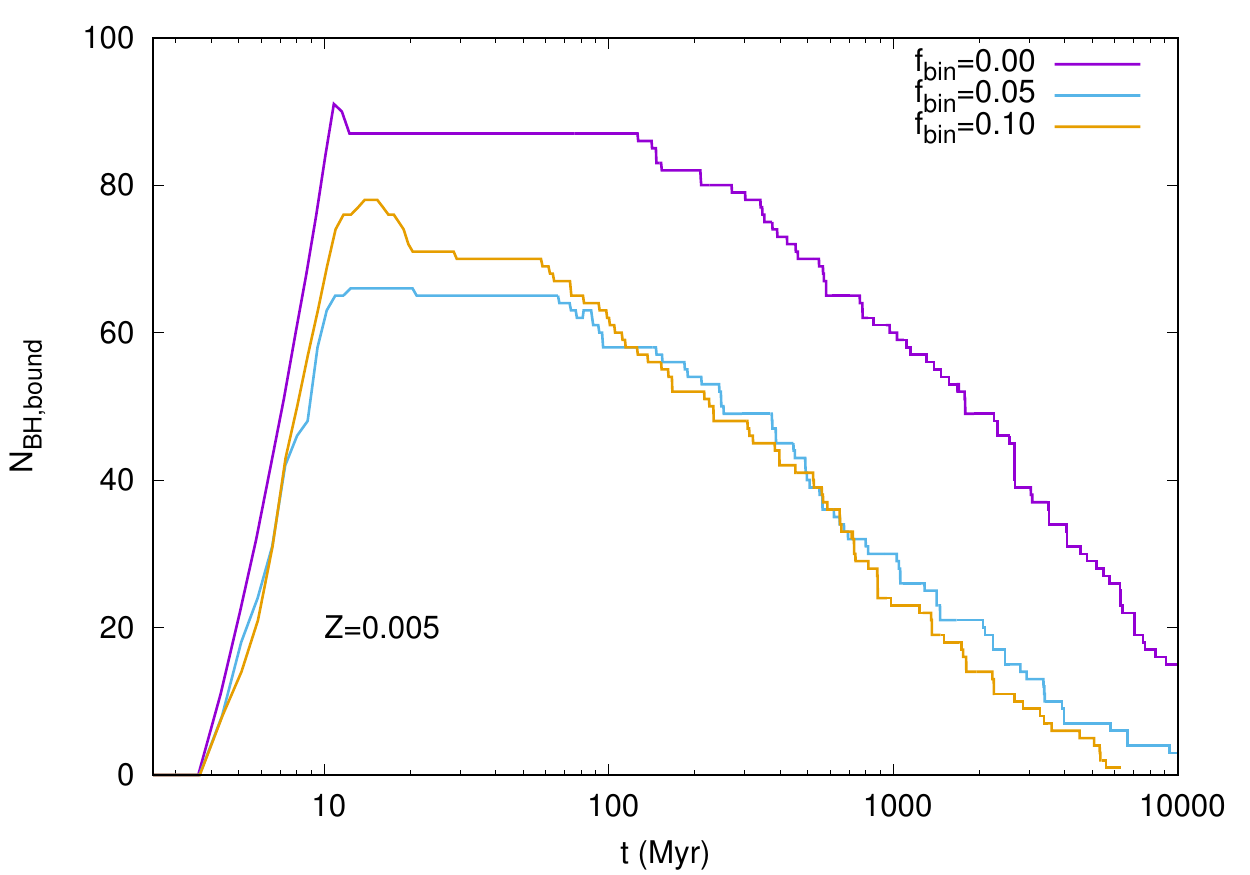}
\caption{The time evolution of the half-mass radii (top panels), the retention fraction
	of BHs (middle), and the total number of bound BHs (bottom) for the
	computed models with initial masses $\mcl(0)\approx3.0\times10^4\Ms$, metallicities
	$Z=0.05\Zs$ (left column) and $0.25\Zs$ (right column), and overall primordial
	binary fractions $\fbin\approx0$\%, 5\% and 10\% (legends). These
	panels demonstrate that increasing the primordial binary fraction
	results in a general increase of the frequency of dynamical scattering interactions
	and energy injection, resulting in a faster decline of the cluster's
	BH population and hence its approach towards core collapse.}
\label{fig:comp_30k}
\end{figure*}

\section{Model calculations: introduction of primordial binaries}\label{calc}

Table~\ref{tab1} summarizes the new computations, as well as those from Paper I.
Extending from Paper I, the newer models here \emph{initiate} with Plummer profiles of masses
$\mcl(0)\approx7.5\times10^4\Ms$ and $1.0\times10^5\Ms$, half-mass radius, $\rh(0)\approx2$ pc, and
with only single stars. Since it takes much longer time to compute
the $\mcl(0)\approx1.0\times10^5\Ms$ models, they are evolved only for $\tevol\approx1$ Gyr, as opposed to
the long-term evolution for the rest of the models, but an additional metallicity of $Z=0.5\Zs$,
on the top of the metallicities $Z=0.25\Zs$ and $0.05\Zs$, are considered for the most massive models.
As in Paper I, the zero-age main sequence (hereafter ZAMS) masses of the individual stars in all the
newly computed models follow the standard \citet{Kroupa_2001} initial mass function (hereafter IMF),
with the maximum stellar mass correlated with $\mcl(0)$ \citep{Weidner_2004,Kroupa_2013}.

In addition to the initially single star-only models, initially Plummer-profiled clusters of
$\mcl(0)\approx3.0\times10^4\Ms$, $\rh(0)\approx2$ pc, $Z=0.25\Zs$ and $0.05\Zs$, and having a primordial-binary
population are evolved for $\approx 10$ Gyr. All YMCs, open clusters,
and GCs are observed to contain a high to a small fraction of stars in
hard (spectroscopic) binaries \citep{PortegiesZwart_2010}. The inclusion of a primordial-binary
population would not only make the models more realistic but would also potentially make the dynamical
interactions richer and versatile, by virtue of the enhanced occasions of exchange
interactions offered by the binaries. That way, they are likely to influence the formation of BBHs
and other types of compact binaries, which effect has not yet been studied through detailed N-body simulations,
especially, for massive and compact systems as in here. 

While the inclusion of primordial binaries is intriguing, it also makes the computations
more rigorous and time consuming, so that an economical strategy is adopted here for the
computations with primordial binaries. First, a high primordial binary fraction of
$\fobin\approx100$\% and the orbital period and eccentricity
distributions of \citet{Sana_2011} are adopted for the O-type stars
(ZAMS mass $\mzams\geq16\Ms$),
to be consistent with the observed high (present-day) binary fractions among O-stars in YMCs,
open clusters, and OB associations \citep{Sana_2011,Sana_2013,Oh_2015}. The binary mass ratios, $q$,
among the O-type stars, follow a uniform distribution (here, an O-star is
paired only with another O-star, as typically observed, and the pairing among the
lower-mass stars is obtained separately; see below).

The binaries among stars with $\mzams<16\Ms$ are obtained separately, where much
lower fractions of primordial binaries are adopted. The primary motivation for this is
to reduce the computing cost; that way a low overall binary fraction can be maintained
throughout the model's evolution, easing the computing demand. For $\mzams<16\Ms$,
primordial binary fractions of $\fbin\approx2$\%, $\approx5$\%, and $\approx10$\%
are taken (see Table \ref{tab1}), that cover the typical range of spectroscopic
binary fractions in GCs and open clusters. Note that since, in all models, the O-type stars comprise
only a small fraction of the total stellar population, the $\fbin$ also
represents the overall primordial binary fraction of the model, as opposed to
the much higher $\fobin\approx100$\% primordial binary fraction among the O-stars themselves.  

The orbital periods of the non-O-star primordial binaries are taken to follow a
\citet{Duq_1991} distribution, and their eccentricities follow
a thermal distribution \citep{1987degc.book.....S},
that characterize a dynamically-processed binary population
\citep{Kroupa_1995a}. Their mass ratios are taken to be uniform between $0.1\leq q\leq 1$.
It would, perhaps, have been more realistic to adopt a ``birth'' population of
primordial binaries \citep{Kroupa_1995a,Kroupa_1995b} or an analytically ``pre-evolved''
(primordial) binary population \citep{Marks_2011}, but they would sustain a much higher
binary fraction for the relatively modest $\mcl(0)\approx3.0\times10^4\Ms$, $\rh(0)\approx2.0$ pc
clusters evolved here, making such long-term computations practically prohibitive.
In that way, the present scheme of including primordial binaries
provides a reasonable compromise between the economy
of computing and consistencies with observations. Most importantly for this study,
the progenitor stars of the BHs (\ie, the O-stars), in the primordial-binary models,
begin in a more realistic environment comprising a high fraction of dynamically-active binaries,
as observed \citep{Sana_2011,Sana_2013}.

In fact, $\fobin\approx100$\% implies
that the consequences of massive binary evolution is automatically taken into account (here, applying
the binary-evolution routine \bse; see below) but in their natural dynamically-active
habitat, given that the majority of the massive stars and their binaries are
observed to reside in dense stellar environments,
as opposed to classical population-synthesis studies of compact-binary formation (see
Sec.~\ref{intro} and references therein) where the massive binaries are evolved without any external perturbation.

No primordial mass segregation is applied to the computed models in Table~\ref{tab1}.

\subsection{The \nbseven N-body evolution program}\label{nbprog}

As in Paper I, all the newer computations here are carried out using the state-of-the-art
direct N-body evolution program
\nbseven\footnote{In practice,
\nbseven and its predecessor \nbsix (see below) are often used synonymously.}
\citep[see also \citealt{Rantala_2017}]{2003gnbs.book.....A,2012MNRAS.422..841A};
see Sec.~2.1 of Paper I for the details on this program. In short, it is a fourth-order
Hermite integrator for tracking the individual trajectories in an arbitrary, self-gravitating,
many-body system where the close encounters and the bound subsystems (binaries and multiples) are treated
with regularization techniques. The integration
is accelerated by utilizing a neighbour-based scheme \citep{Nitadori_2012} for the force contributions
at the shortest time intervals (the ``irregular'' force/steps).
At longer time intervals (the ``regular'' force/steps), all members in the
system are included for the force evaluation. The irregular forces
are computed by parallel processing in CPUs, while the much more
expensive regular force evaluations are done on 
GPUs\footnote{All computations in this work are done on
workstations equipped with quad-core {\tt AMD} processors and {\tt NVIDIA}'s
{\tt Fermi} and {\tt Kepler} series GPUs.}. The diverging gravitational forces
during close passages and in binaries are dealt with two-body or KS regularization
\citep{2003gnbs.book.....A} and higher-order multiples are treated with the
Algorithmic Regularization Chain
(ARC; \citealt{Mikkola_1999,Mikkola_2002,Mikkola_2008}). Unlike
the traditional- or KS-Chain Regularization \citep{Mikkola_1993}, that is utilized in
\nbsix, the application of the ARC in \nbseven allows the inclusion of members, in the Chain,
with arbitrary mass ratios. This makes \nbseven more suitable for the
present calculations where the BHs are of a wide mass range and predominantly
take part in binary-single and binary-binary interactions (see Sec.~\ref{intro},
below).

In \nbseven, the Chain members and the Chain perturbers are, however,
selected in the same well-proven way as in the KS-Chain implementation
in \nbsix \citep{2003gnbs.book.....A,2012MNRAS.422..841A}, except that the internal integration
of the Chain
is done in the ARC way (\citealt{Mikkola_2008} and references therein). Here,
the Chain members, that comprise a compact subsystem
(often a triple or a quadruple), are initially selected based on the
strategies for identifying hierarchical systems. This comprises
identifying, for each particle in the system, the dominant and the next-to-dominant
perturber and testing whether to form a
triple or a quadruple; see Chapter 9 and Algorithm 11.3 of \citet{2003gnbs.book.....A}
for the details. The
Chain is then constructed (and possibly switched during the Chain integration
due to either change in the subsystem's configuration or
the close approach of a perturber) in the usual way
by sequentially connecting the closest subsystem members with vectors as described in
\citet{Mikkola_1993}. The perturbers of the compact subsystem are selected
either from within its close-encounter radius or otherwise through a full perturber
search.
The key algorithms are discussed in Chapter 12 of \citet{2003gnbs.book.....A}.

An important aspect of \nbseven is its GR treatment of binaries and multiples
containing a BH or a neutron star (hereafter NS), through the ARC
\footnote{To accommodate the unlikely event of a pure KS binary becoming relativistic,
GR orbital modifications are included in the KS treatment as well.}
\citep{Mikkola_2008}. This allows for
on-the-fly GR orbital modifications and coalescences of relativistic subsystems (typically
a binary or a triple containing one or more BH/NS) that are bound to the system.
In principle, PN-1, PN-2 (GR periastron precession), PN-2.5 (orbital shrinking due to
GW radiation), PN-3, and PN-3.5 order terms can be included in the ARC
procedure, including their spin contributions.
For the ease of computing, the PN orders are increased sequentially as they become
significant; see \citet{2012MNRAS.422..841A} and references therein for the details of the implementation
and \citet{Brem_2013} for an alternative approach (in \nbpp).
However, for the
economy in computing time, the BHs' spins are taken to be zero in the present computations.
The spin terms would have modified the times of GR coalescences occurring within the cluster
(see Secs.~\ref{intro} \& \ref{bbh}) to some extent, however, this is
not critical here due to the statistical nature of the dynamically-induced
BBH coalescences.
The latest implementation shows reasonable energy check
(typical relative energy change $\sim10^{-4}-10^{-6}$) even during extreme relativistic
events such as a BBH coalescence within a triple.

In reality when the BHs have spins,
a BBH would typically receive a large GW merger kick during its
inspiral phase ($\sim100-1000\kmps$; \citealt{Campanelli_2007,Hughes_2009}). 
This would cause the newly-formed merged BH to escape from the cluster almost immediately,
and it would hardly have a chance to participate in dynamical encounters further.
This situation is mimicked by applying a velocity kick onto the merged BH
immediately after a GR BBH
coalescence occurs within the cluster (Sec.~\ref{bbh}).
In the current implementation in \nbseven, the applied kick
is kept only marginally above the escape speed to avoid large energy errors;
$\approx5$ times the central RMS speed.
This is still enough to eject the merged BH out of the cluster in several dynamical times;
in reality a BBH coalescence product would typically escape at a much higher
speed. In any case, it is found in test calculations that even if a merged BH is retained,
it does not necessarily participate in further GR coalescences.

\nbseven utilizes the semi-analytic single- and binary-stellar evolution algorithm
{\bse} \citep{Hurley_2000,Hurley_2002}
to evolve each star/binary and form their remnants (see Sec.~2.1 of Paper I). For comparison purposes,
the same recipes of stellar evolution and remnant formation as in Paper I (see Sec.~2.2 of paper I)
are adopted here for the newer computations. In brief, this involves a modified  
version of the \bse that adopts
the state-of-the-art stellar-wind and remnant-formation schemes of
\citet{Belczynski_2008,Belczynski_2010}, thereby allowing the formation of DCBHs
and of NSs via electron-capture supernovae (ECS; \citealt{Podsiadlowski_2004}),
from both single stars and binary members. The DCBHs and the ECS-NSs are the massive-stellar
remnants that receive zero natal kicks and are retained in the cluster at birth.

It is to be noted that the single-stellar BH mass spectrum (see Fig.~1 of Paper I
and Fig.~\ref{fig:mdist_BH} here)
gets modified when nearly all the BH progenitors are in binaries, as in the
primordial-binary models here. This is due to the mergers of
massive binaries or of binary
components during binary-binary encounters, resulting in more massive progenitors and hence
more massive BHs, and also due to the mass transfer, CE, and tidally-enhanced
stellar winds that might occur within the binaries, modifying the components' mass
evolution. The BH masses are likely to be as well different from those in a pure
population synthesis approach (for given stellar- and binary-evolution schemes;
here the modified \bse) since the stochastic dynamical perturbations and close interactions
would modify the binaries' evolutions. The mergers among the massive primordial binaries would
also cause less number of DCBHs (but with a broader mass spectrum) to be produced compared
to the initially single-only counterpart; see the bottom panels of Fig.~\ref{fig:comp_30k}.
Fig.~\ref{fig:mdist_BH} demonstrates the difference in the BH mass spectra in the
two cases for the $\mcl(0)\approx3.0\times10^4\Ms$ models;
the distributions here correspond to an early evolutionary time
of $t\approx20$ Myr and include the non-DCBHs too, that would escape soon after due to their natal kicks,
although there are only a handful of such BHs per cluster for the $Z\leq0.25\Zs$
considered here.

\begin{figure*}
\centering
\includegraphics[width=8.5cm,angle=0]{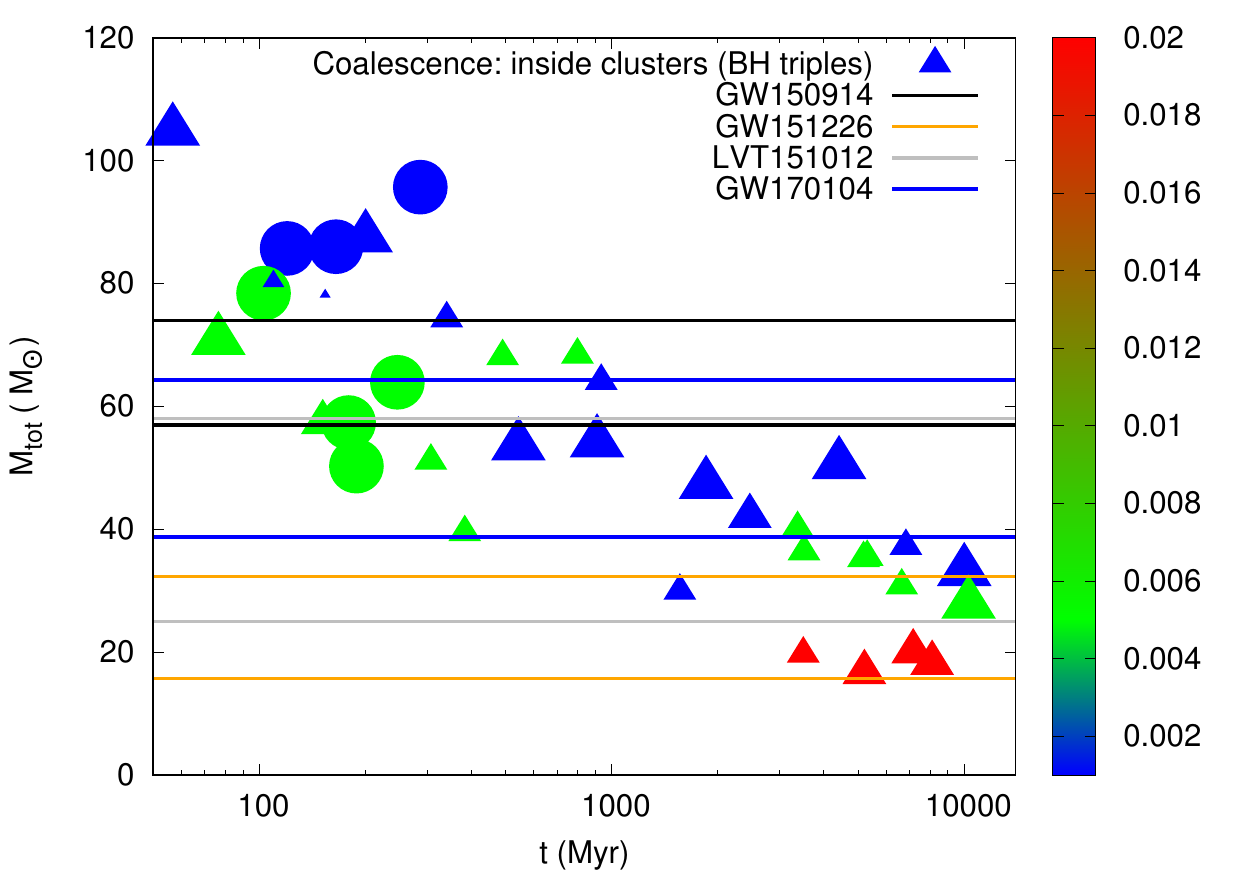}
\includegraphics[width=8.5cm,angle=0]{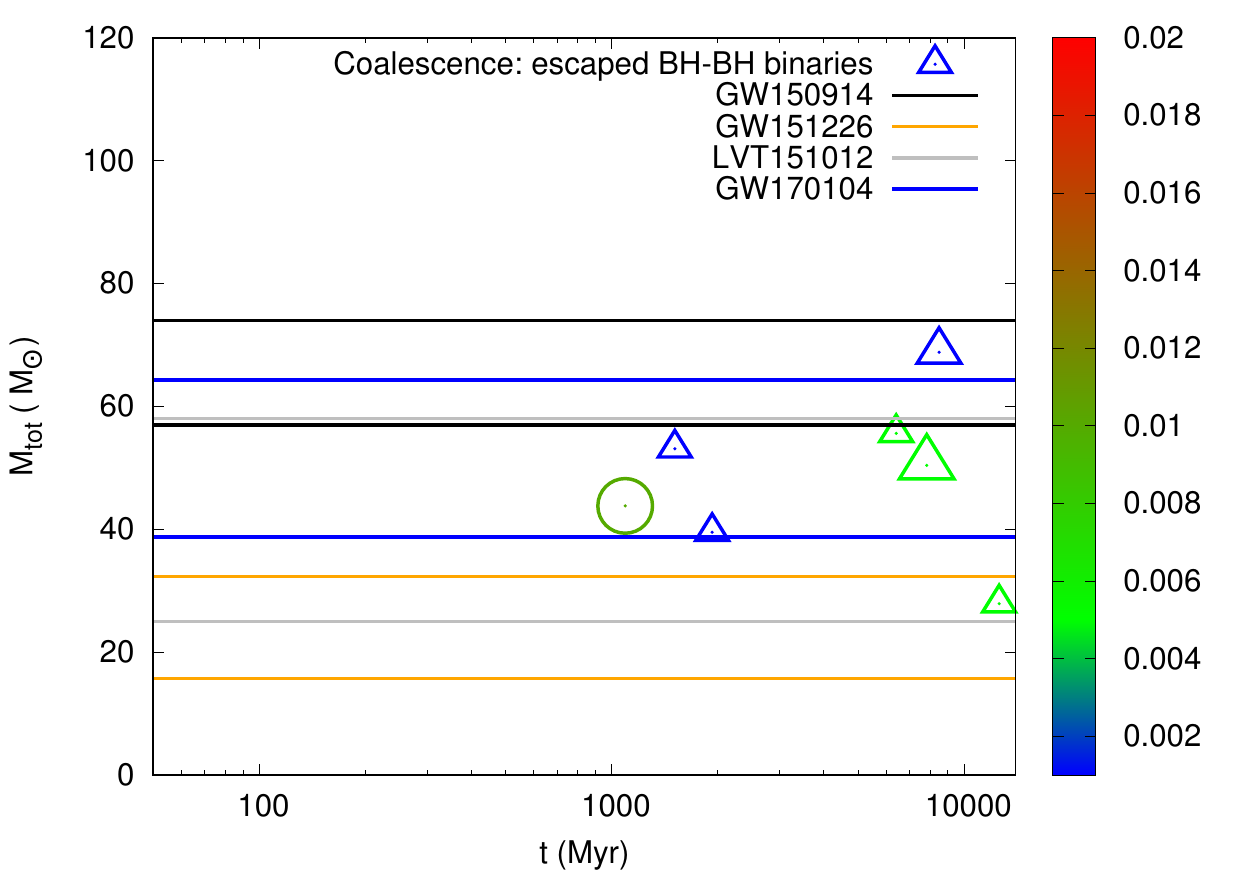}
\includegraphics[width=8.5cm,angle=0]{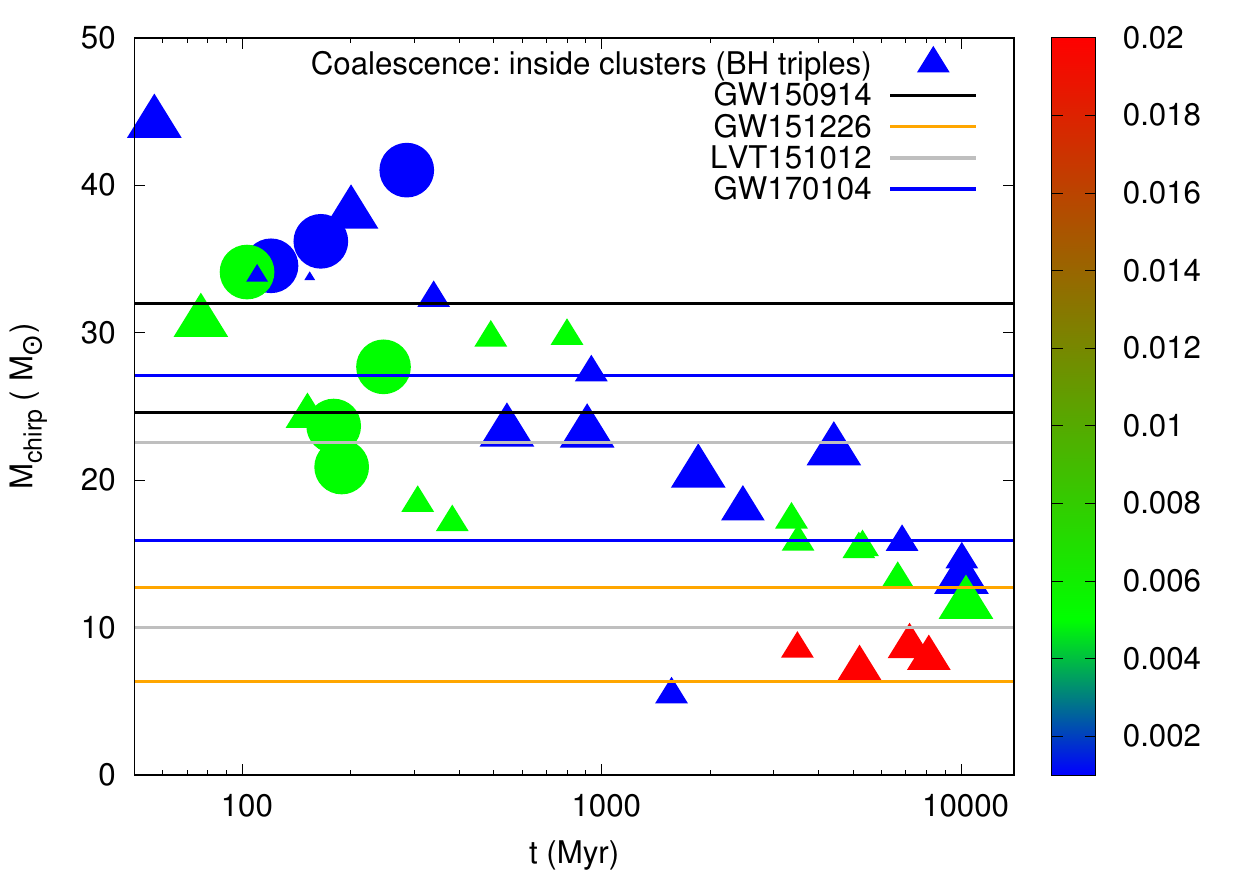}
\includegraphics[width=8.5cm,angle=0]{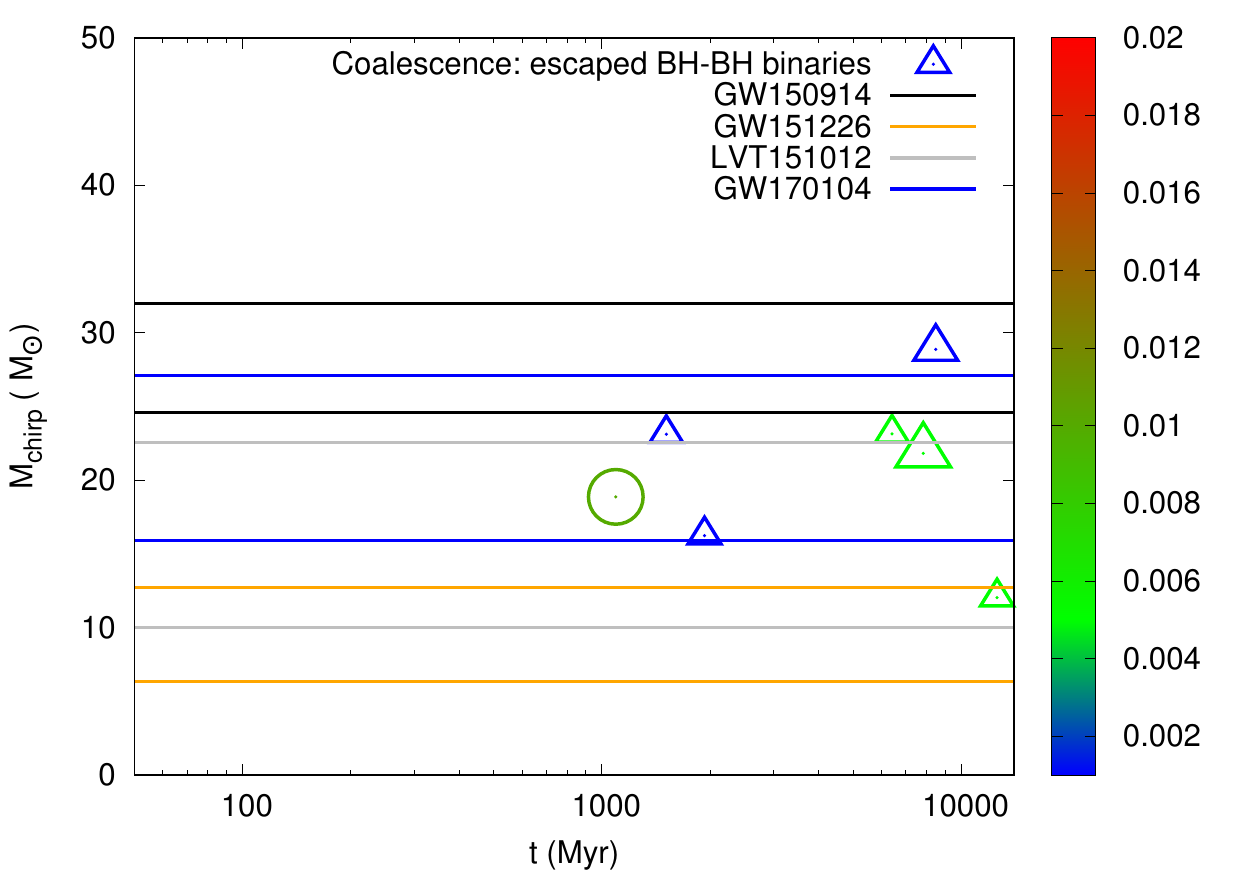}
\includegraphics[width=8.5cm,angle=0]{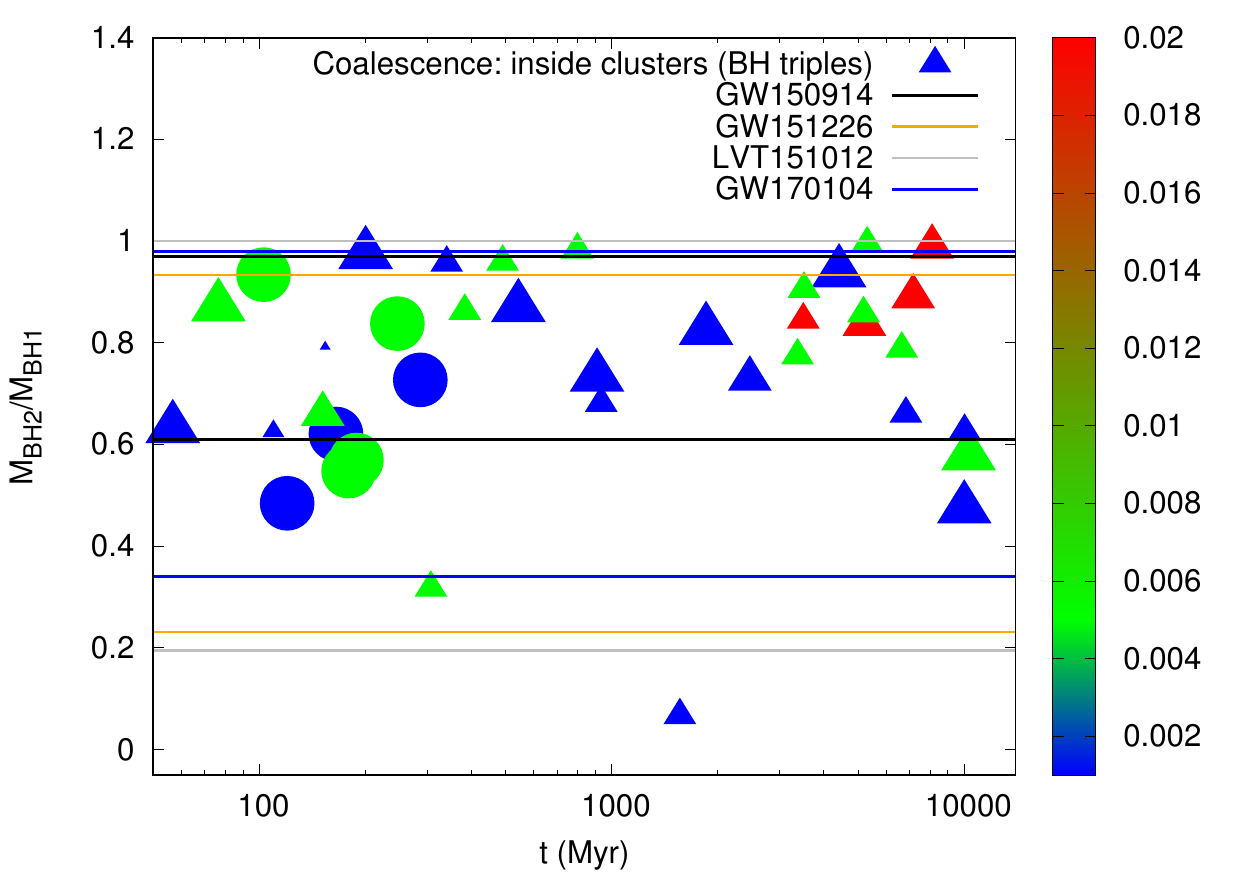}
\includegraphics[width=8.5cm,angle=0]{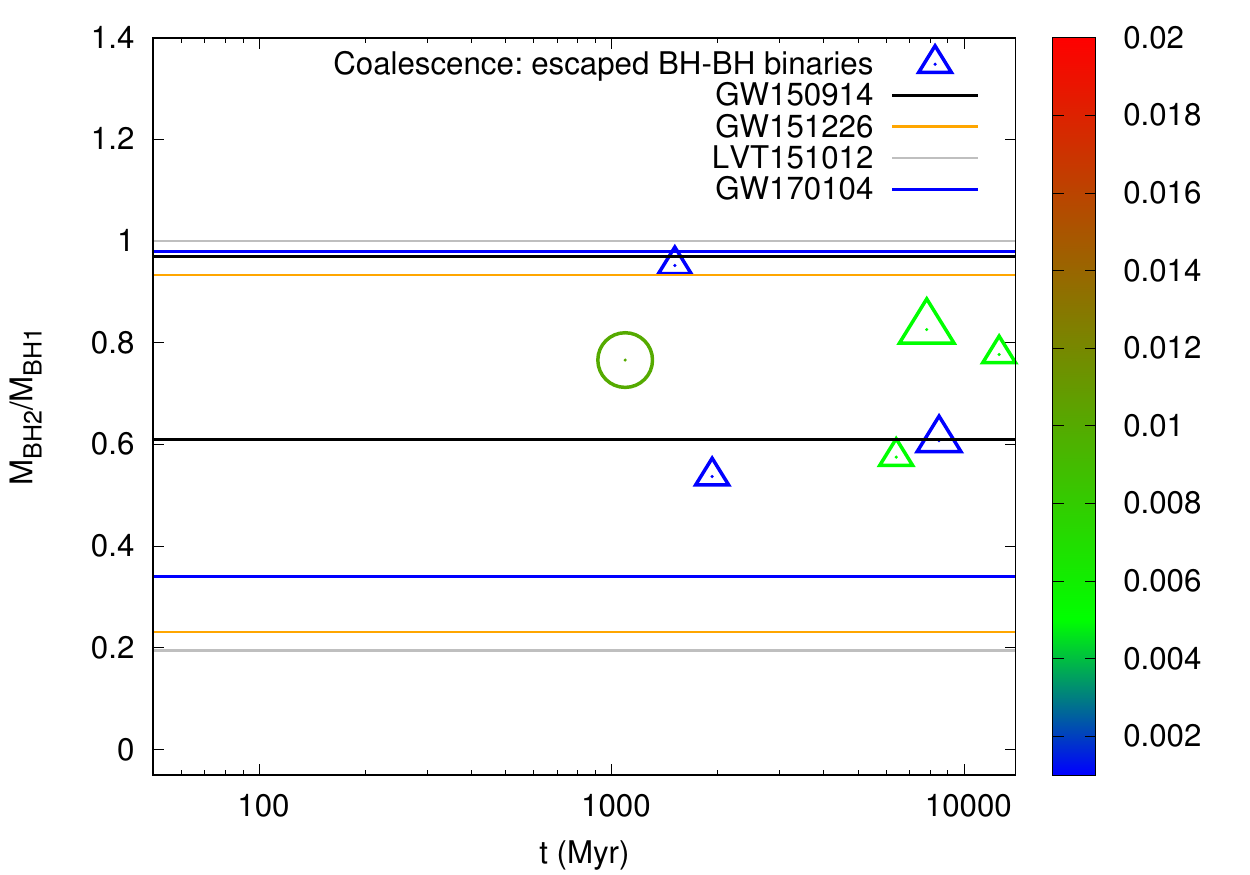}
	\caption{{\bf Left column:} The total mass, $\mtot$ (top panel), chirp mass, $\mchirp$ (middle),
	and the mass ratio, $M_{\rm BH2}/M_{\rm BH1}$ ($M_{\rm BH2} \leq M_{\rm BH1}$; bottom),
	of the (triple-mediated) BBH coalescences that occur while being bound to the
	clusters, against the cluster-evolutionary times, $t=\tmrg$, at which the mergers
	occur. The mergers for all models with initial masses $\mcl(0)\lesssim7.5\times10^4\Ms$
	(see Table~\ref{tab1})
	are represented by filled triangles whose sizes are scaled in proportion to the clusters'
	$\mcl(0)$s and which are colour coded according to the models' assumed metallicities (the
	colour bars; Table~\ref{tab1}).
	The mergers within the $\mcl(0)\approx10^5\Ms$ models are denoted by filled
	circles which have the same colour coding. The particular coalescence
	with mass ratio $\approx0.1$ (see bottom panel) is actually an NS-BH merger.
	{\bf Right column:} The same as the left column but for the coalescences, occurring
	within the Hubble time,
        among the BBHs that are ejected from the clusters and for the mergers being represented
	by the corresponding empty symbols. Here, the plotted merger times, $t=\tmrg$, include the
	cluster-evolutionary times, \tej, of the BBHs' ejections, \ie, $\tmrg\equiv\tej+\taumrg$,
	where $\taumrg$ is the GW merger time \citep{Peters_1964} of the BBH right after its ejection.   
	}
\label{fig:bbhmrg}
\end{figure*}

\section{Results}\label{res}

Fig.~\ref{fig:comp_all} compares the time evolution of the half-mass radius, $\rh(t)$, and
of the fraction of the bound BHs, for the initially single star-only models in Table~\ref{tab1}.
As already noted in Paper I (see its Sec.~3.1),
for a given initial size $\rh(0)\approx2$ pc and a given $Z$
(BH mass spectrum), the dynamical heating of the BHs begins to expand an
initially less massive (dense) cluster at a somewhat larger rate. However,
in the long run, more numerous BHs in a higher $\mcl(0)$ system continue
to expand their host for a longer time, that, in turn, inhibits the dynamical
self depletion of the BHs, ultimately continuing to retain a larger
fraction of the initially-bound BHs (\cf Fig.~5 of Paper I). As pointed out in Sec.~\ref{intro},
such self-regulation is responsible for the long-term retention of BH populations, especially in
massive low-$Z$ systems such as GCs, where stellar-mass BH candidates have recently been
identified.

Fig.~\ref{fig:comp_30k} compares the evolution of the size and the BH retention fraction
among the $\mcl(0)\approx3.0\times10^4\Ms$ models with and without primordial binaries (see
Table~\ref{tab1}). Interestingly, the BHs deplete faster, in long term, in the presence of
a higher primordial binary fraction (Fig.~\ref{fig:comp_30k}, middle panels).
This is a consequence of the overall increase in the rate of strong
scattering encounters, with increasing primordial binary fraction (among the
lower-mass stars; see Sec.~\ref{calc}), that the BHs experience while they continue to
interact with the other stellar members of the cluster (see Sec.~\ref{intro}). The correspondingly
enhanced rate of BH loss, in turn, stalls the expansion of the cluster and resumes
its secular contraction \citep{1987degc.book.....S} earlier (Fig.~\ref{fig:comp_30k},
top panels). The presence of primordial binaries, therefore, might be the key
to prevent old, low-$Z$ GCs to grow to unusually large sizes
(\cf, \citealt{Morscher_2015,2016arXiv160300884C}).

\subsection{Dynamically-formed BBH: gravitational-wave coalescence events}\label{bbh}

The sixth (seventh) column of Table~\ref{tab1} enlists the individual BBH coalescences that occur
while being bound to the cluster (occur within a Hubble time after being ejected from the cluster),
where the merging masses ($\mbhone + \mbhtwo$) and the corresponding cluster-evolutionary
times\footnote{Unless otherwise stated, all merger times, \tmrg, are measured
w.r.t. the beginning of the parent cluster's evolution.}
for the merger (ejection), \tmrg(\tej)\footnote{The instant of ejection of any entity is taken as
the time at which it crosses the tidal radius.},
are provided in the parentheses. As already seen in Paper I,
the BBH coalescences\footnote{In \nbseven, the speed of light can be ``chosen''
so as to make the system more or less relativistic. In all the computations
presented here, the physical speed of light is always chosen.},
in these models, occur predominantly while being bound to the cluster,
rather than happening among those BBHs that are ejected from the cluster. As discussed
in Sec.~3.2 of Paper I, this is expected for the lower-mass, open-type clusters, as simulated
here, in the presence of a broad BH mass spectrum, as obtained from the present remnant-formation
schemes. Such in-cluster or in situ BBH mergers are typically induced via large eccentricity boost
of the merging BBH, while it is
a part of a triple (with typically a BH or possibly with a normal stellar member), due
to the triple's internal dynamical evolution.
Such triples continue to form and disassemble due to the numerous close
encounters involving single BHs, BBHs and/or normal stellar binaries
(see Sec.~\ref{intro} and references therein).

In fact, as apparent from Table~\ref{tab1}, the number of such triple-mediated BBH mergers, in general,
increase dramatically in the presence of primordial binaries; \cf the $\mcl(0)\approx3.0\times10^4\Ms$
models in Table~\ref{tab1} with $\fbin=0$ and $\fbin>0$. Like the enhanced depletion
rate of the BHs (see above; Fig.~\ref{fig:comp_30k}), this is also a consequence of the persistent
population of (lower-mass) primordial binaries being continuously engaged in close encounters  
with the BHs, thus allowing additional BBH formation via exchange encounters.
An increase of the number of in situ BBH mergers, with the overall binary fraction, can also be noted in
the much more massive clusters in Monte Carlo-based studies, \eg, in \citet{Morscher_2015}. 
In fact, all the in situ BBH coalescences in the primordial-binary models here involve BHs from
different primordial binaries, \ie, all of them are assembled through exchange interactions.
For the $\fbin>0.05$ clusters, 10-20\% of the \emph{ejected} BBHs, per model, contain BH members
that are derived from the same primordial binary.

As an inspection of the current computation outputs reveals, the majority of the BBH
inspirals occur in triples that are intermediate or metastable in nature (Sec.~\ref{intro}
and references therein). However, although rare, cases are spotted where a highly hierarchical
and long-lasting triple has ultimately led to the GR inspiral and merger of its inner BBH through the
eccentric Kozai mechanism, in its regular sense (Sec.~\ref{intro} and references therein).
Note that hierarchical triples, in dense systems
such as here, are often perturbed by intruders
so that their evolutionary and stability properties differ from those had they evolved
in isolation. A more thorough study of the nature of the (resonant/hierarchical) triples/multiples hosting
a BBH merger, in realistic cluster environments as here, is currently ongoing (Banerjee,
in preparation). In the rest of this paper, the in-cluster GR mergers will simply be
denoted as ``triple-induced mergers''.

In contrast, Monte Carlo simulations of much more massive, GC-like clusters,
typically of $10^5\Ms\lesssim\mcl(0)\lesssim10^6\Ms$ (see Sec.~\ref{intro}
and references therein),
predominantly yield ejected BBH mergers. Such massive systems, typically unapproachable
by the direct N-body method (but see \citealt{Wang_2016}; Sec.~\ref{intro}),
eject BBHs that are much tighter, owing to the clusters' large escape speeds, facilitating mergers
among them, unlike those here. It should, however, be noted that a direct N-body
approach, with explicit GR treatment, deals with subsystems more
consistently and completely than in a typical Monte Carlo approach (see Sec.~\ref{intro} and references therein);
the former approach would perhaps inherently
(and more realistically) lead to a larger number of on-the-fly, triple-induced GR coalescences
than its Monte Carlo counterpart. At present, the ranges of models computed
with the two techniques by different groups and the details of the dynamical treatments adopted in them 
barely overlap. This makes a direct comparison among the two
regimes and the methods, especially in terms of rare strong-encounter products
such as dynamical BBH mergers, difficult. However, in terms of the overall long-term behaviour
of the clusters, Monte Carlo and direct N-body approaches have shown to
yield mutually agreeable outcomes (see, \eg, \citealt{Giersz_2013,Rodriguez_2016c}).

The left (right) column of Fig.~\ref{fig:bbhmrg} shows the triple-induced (ejected) 
BBH mergers' total mass, $\mtot$, chirp mass, $\mchirp$ \footnote{The chirp
mass between two masses $m_1$ and $m_2$ is defined as
$\mchirp\equiv\frac{(m_1m_2)^{3/5}}{(m_1+m_2)^{1/5}}$.},
and mass ratio,
$\mbhtwo/\mbhone{\rm~}(\mbhtwo<\mbhone)$, against their
respective merger times, $t=\tmrg$, where the outcomes of all
the models in Table~\ref{tab1} are superimposed. A clear negative trend is visible between the 
\emph{bound} BBH mergers' $\mtot$ or $\mchirp$ and their $\tmrg$s
(Fig.~\ref{fig:bbhmrg}, left column): this is simply due to
the fact that the more massive BHs generally remain more centrally concentrated and hence
predominantly get engaged in dynamical interactions, at earlier evolutionary times.
Such a behaviour is also seen in the merger times of \emph{ejected} BBHs from massive GCs;  
see, \eg, \citet{2016arXiv160906689C}.

On the other hand, in the present computations, no clear trend between the $\tmrg$s and the
masses is apparent for the ejected BBHs (Fig.~\ref{fig:bbhmrg}, right column). First,
there are only a few ejected BBHs with $\tmrg<13.7$ Gyr.  
A trend is furthermore washed out due to the fact that the occurrence of a merger within a Hubble
time, for these relatively wide (${\rm semi-major-axis}=a$), ejected BBHs,
relies mostly on their dynamically-induced, very high (initial) eccentricities, $e$,
which sensitively controls the binaries' GR coalescence time, \taumrg, through \citep{Peters_1964}

\begin{equation}
\taumrg\approx\frac{5}{64}\frac{c^5a^4(1-e^2)^{7/2}}
	{G^3 m_1 m_2 (m_1+m_2)}
	\left(1+\frac{73}{24}e^2+\frac{37}{96}e^4\right)^{-1},
\label{eq:taumrg}
\end{equation}
where $m_1=\mbhone$ and $m_2=\mbhtwo$ are the members' masses of
the BBH\footnote{Note that Eqn.~\ref{eq:taumrg}, which is used in plotting
the data points in the right panels of Fig.~\ref{fig:bbhmrg}, is an approximate formula that,
depending on the initial $e$, underestimates $\taumrg$. A more accurate value of
$\taumrg$ can be obtained by explicitly integrating the \citet{Peters_1964}
expressions of (orbit-averaged) $a$ and $e$ decay, as done in Sec.~\ref{inspiral}.
The sensitive dependence of the \taumrg of the inspiral on the initial $e$ remains as well valid
in the more accurate treatment.}.

\begin{figure*}
\centering
\includegraphics[width=8.5cm,angle=0]{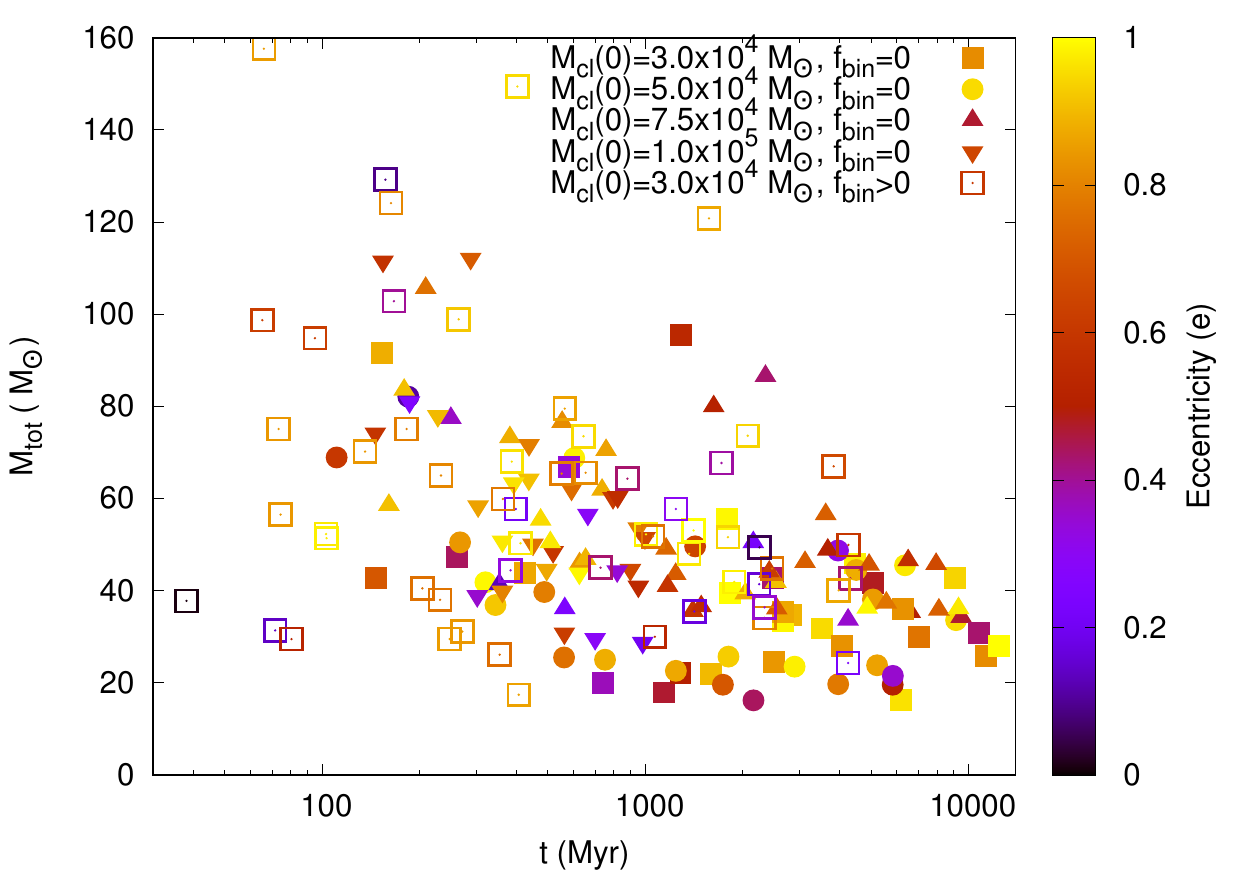}
\includegraphics[width=8.5cm,angle=0]{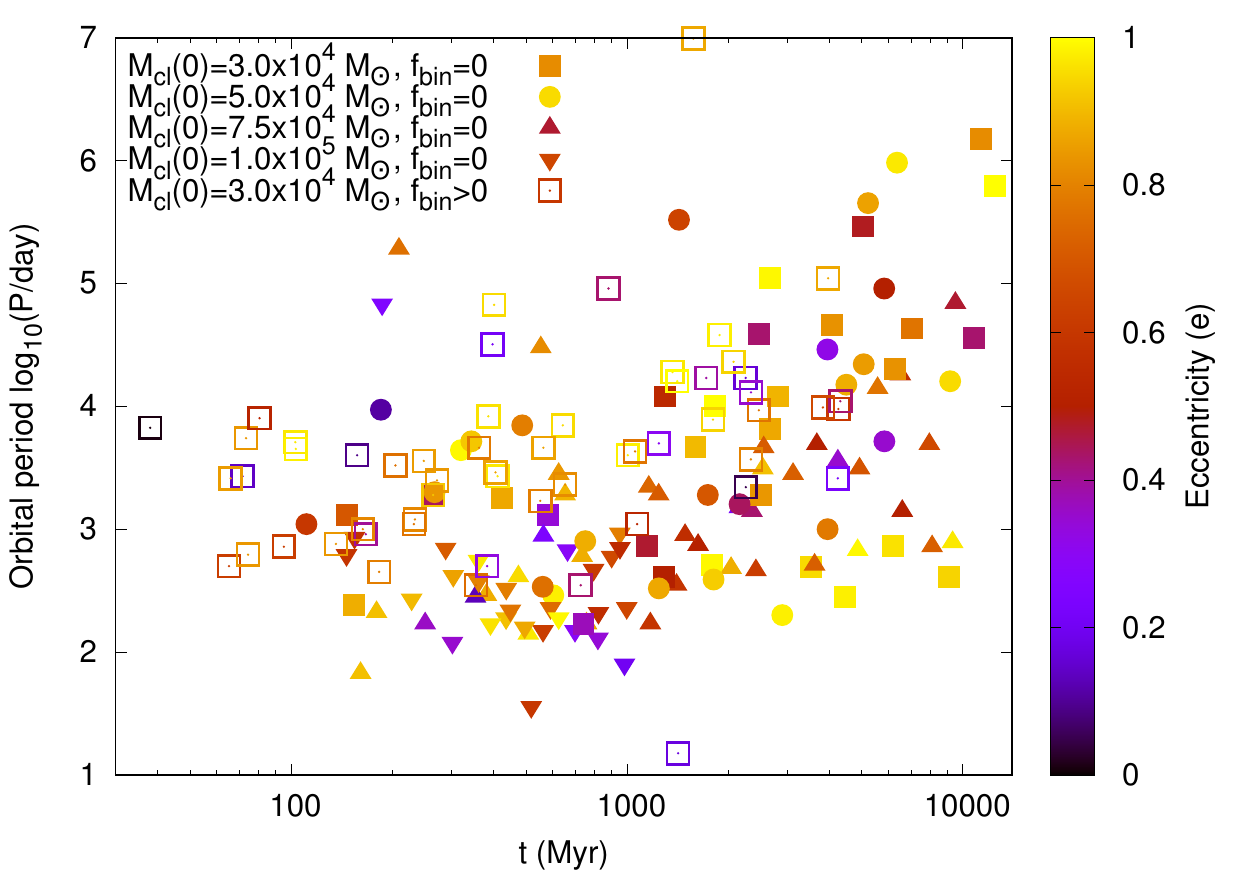}
	\caption{The ejected BBHs from all the $\mcl(0)\geq3.0\times10^4\Ms$ models
	in Table~\ref{tab1}. The BBH data points are distinguished w.r.t. their parent
	clusters' $\mcl(0)$ and primordial-binary content (legends) and are colour-coded w.r.t. their eccentricities, $e$,
	at the instant of ejection (colour bars). The total masses, $\mtot$s, of the BBHs show
	an overall decreasing trend w.r.t. their cluster-evolutionary times, $t=\tej$,
	of ejection (left panel) and their orbital periods, $P$, at the instant of ejection,
	show an increasing trend (right panel). These trends qualitatively conform to what
	can be expected from the secular dynamical evolution of the clusters; see text.}
\label{fig:bbhesc}
\end{figure*}

Fig.~\ref{fig:bbhesc} (left panel) demonstrates the overall negative trend of the
$\mtot$s of the ejected BBHs w.r.t. their $\tej$s, as for the in situ BBH coalescences
(see above; Fig.~\ref{fig:bbhmrg}).
The formation of higher-mass BHs in the models
with primordial binaries (see Sec.~\ref{nbprog}; Fig.~\ref{fig:mdist_BH}) results
in the ejection of more massive BBHs from such models, especially at early evolutionary times.
Higher $\mcl(0)$ models generally tend to eject tighter BBHs (with shorter orbital periods),
owing to larger escape speeds from their centers. The orbital periods, $P$, of the ejected
BBHs have an overall positive correlation (despite large scatter) with $\tej$: this is
due to the expansion and mass loss of the clusters with time (see above), diluting
their central potentials and correspondingly reducing their escape speeds. These
trends are demonstrated in Fig.~\ref{fig:bbhesc} (right panel;
see also \citealt{2016arXiv160300884C}). Note that the ejected BBHs with the
smallest $P$s are generally moderately eccentric (see colour coding in Fig.~\ref{fig:bbhesc}),
so that such systems with $\taumrg<$ 13.7 Gyr are rare, for the cluster mass range explored
here. Also, due to the enhanced rate of scattering and exchange encounters (see above),
the ejected BBHs from the $\fbin>0$ models are more numerous compared to those from their $\fbin=0$
counterparts (open squares in Fig.~\ref{fig:bbhesc}).

\begin{figure*}
\centering
\includegraphics[width=10.0cm,angle=0]{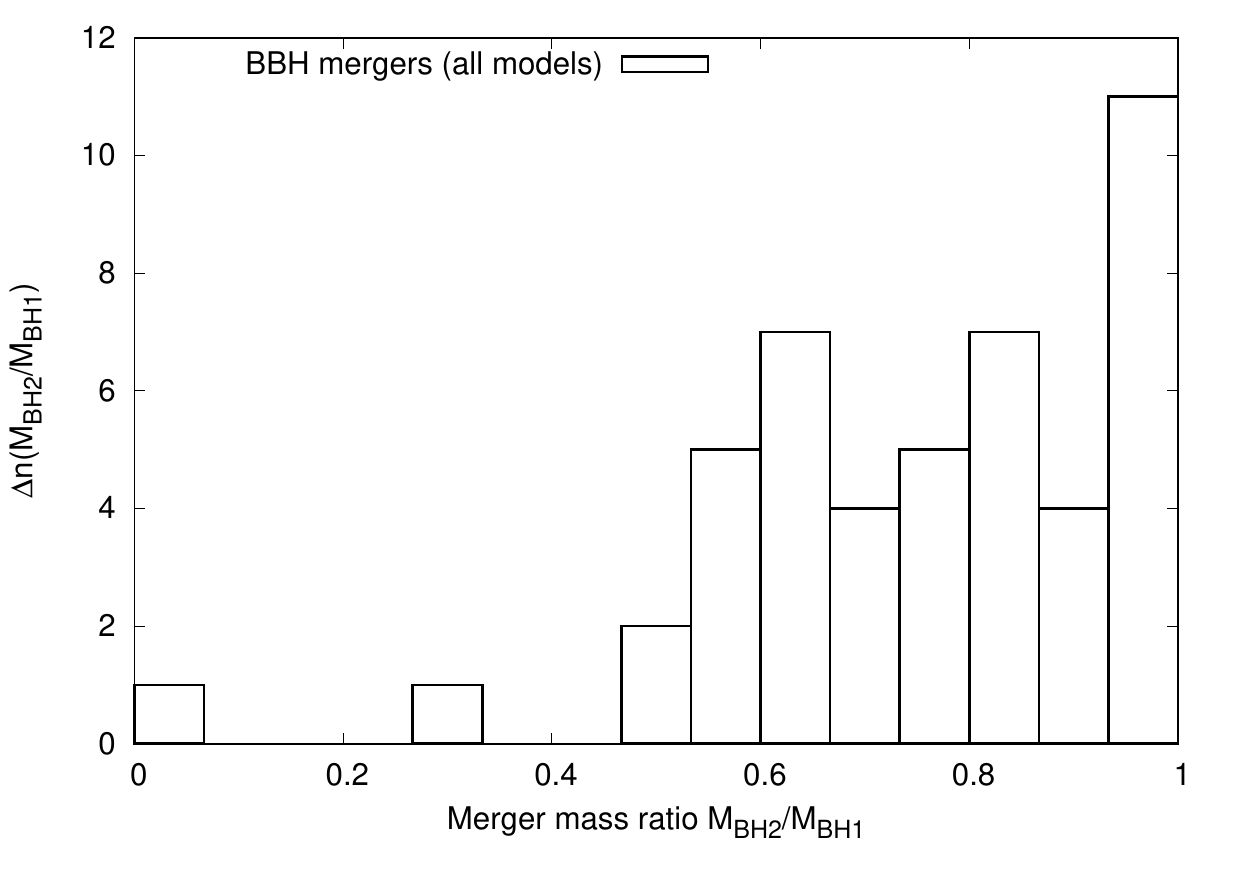}
	\caption{The distribution of mass ratios, $\mbhtwo/\mbhone$, of the BBH mergers
	from all of the models in Table~\ref{tab1} combined (includes both the in situ and
	escaped mergers). The distribution is approximately flat for $\mbhtwo/\mbhone\gtrsim0.5$
	with a peak at $\mbhtwo/\mbhone\approx1$. The merger with the lowest mass ratio is
	an NS-BH merger (see Fig.~\ref{fig:bbhmrg}, Sec.~\ref{nsseg}).}
\label{fig:mratio}
\end{figure*}

As hinted in Paper I, although dynamical selection tends to pair BHs of similar
masses (see Sec.~3.2 of Paper I), the mass ratios of the BBH mergers (both the bound and the
ejected ones) would still have a rather broad range,
typically $0.5\lesssim \mbhtwo/\mbhone \lesssim 1.0$. This is demonstrated
in Fig.~\ref{fig:bbhmrg} (bottom panels); the one with the mass ratio $\approx0.1$ is actually
an NS-BH merger (see Sec.~\ref{nsseg}). Fig.~\ref{fig:mratio} demonstrates that
the distribution of the mass ratios of the BBH mergers, for $\mbhtwo/\mbhone\gtrsim0.5$,  
is approximately flat except for the spike at $\mbhtwo/\mbhone\approx1.0$.
On the other hand, the BBH formation and coalescence
channels involving purely massive binary evolution tend to
produce BBHs and their mergers with mass ratios, typically,
$\mbhtwo/\mbhone\gtrsim0.7$ (\eg, \citealt{Marchant_2016,Belczynski_2016}).
However, \citet{2017arXiv170607053B} have recently produced a wider range of BBH-merger mass
ratio, with modified input physics.   
Hence, a mass ratio substantially less than unity (as in GW151226 and GW170104 events)
can be taken as an indication of the involvement of dynamical encounters in assembling a BBH merger,
besides, any spin-orbit misalignment (see Sec.~\ref{intro}).

The limits of $\mtot$, $\mchirp$, and $\mbhtwo/\mbhone$, for the events detected so far
by the LIGO, are indicated in the panels of Fig.~\ref{fig:bbhmrg}, where the data-points
generally conform with these limits.
Among the triple-induced (in situ) mergers occurring within
a few 100 Myr of the parent cluster's evolution (Fig.~\ref{fig:bbhmrg},
left panels), there are ones that are significantly more massive than even
the most massive detected merger, namely GW150914. In terms of $\mtot$ or $\mchirp$,
the computed clusters typically generate GW150914-like and super-GW150914
mergers within them during their YMC phase ($t\sim10-100$ Myr), GW150914-like, GW170104-like,
and GW151226-like mergers during their intermediate ages ($t\sim$ Gyr),
and only GW151226-like mergers over their latest ages. LVT151012-like mergers are produced
within the clusters from intermediate age onwards.
Note that GW170104-like mergers are produced (within the low-$Z$
clusters) over the widest age range of the host clusters, from the YMC phase until nearly 10 Gyr age.
In all the models
computed here, all types of \emph{ejected} BBH mergers take place
from the clusters' intermediate age onwards (Fig.~\ref{fig:bbhmrg},
right panels).

\begin{figure*}
\centering
\includegraphics[width=13.0cm,angle=0]{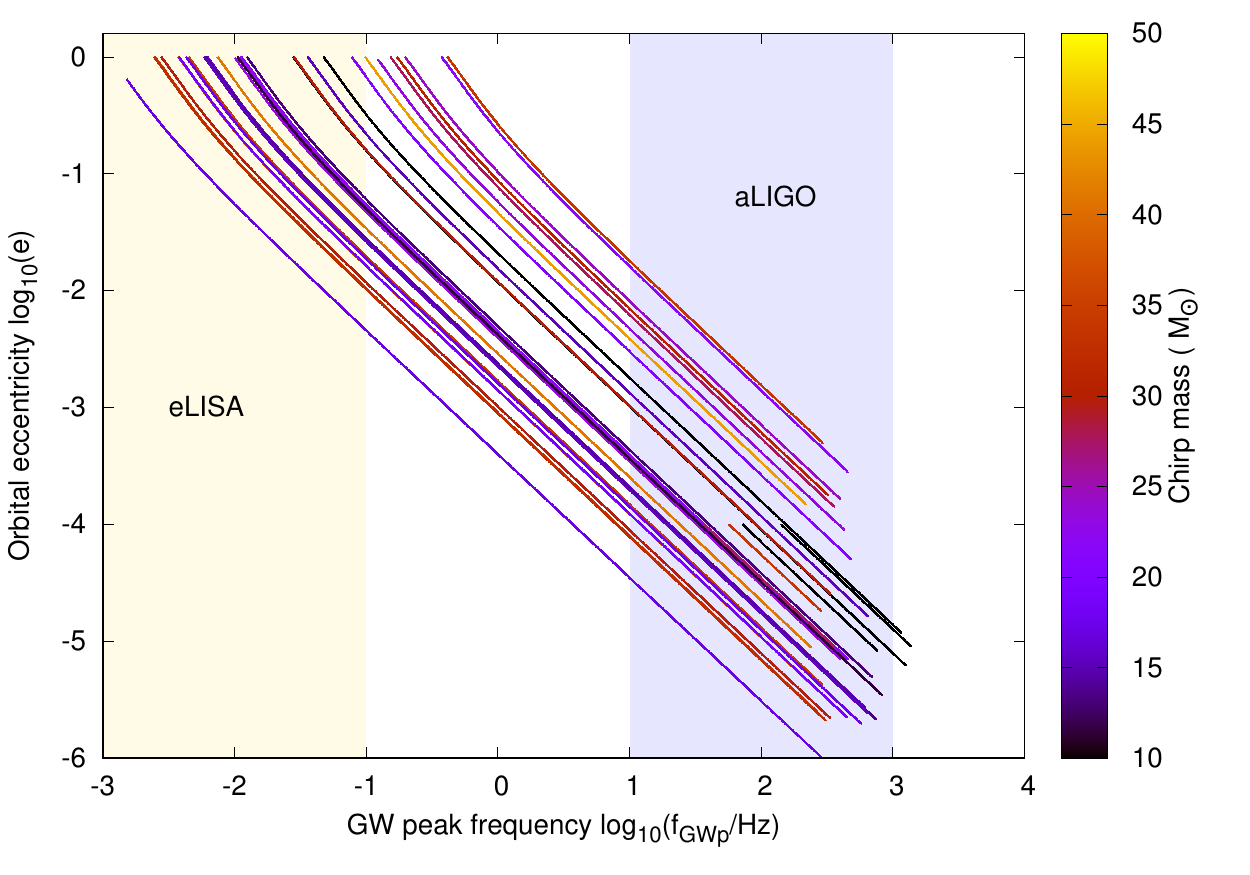}
\includegraphics[width=13.0cm,angle=0]{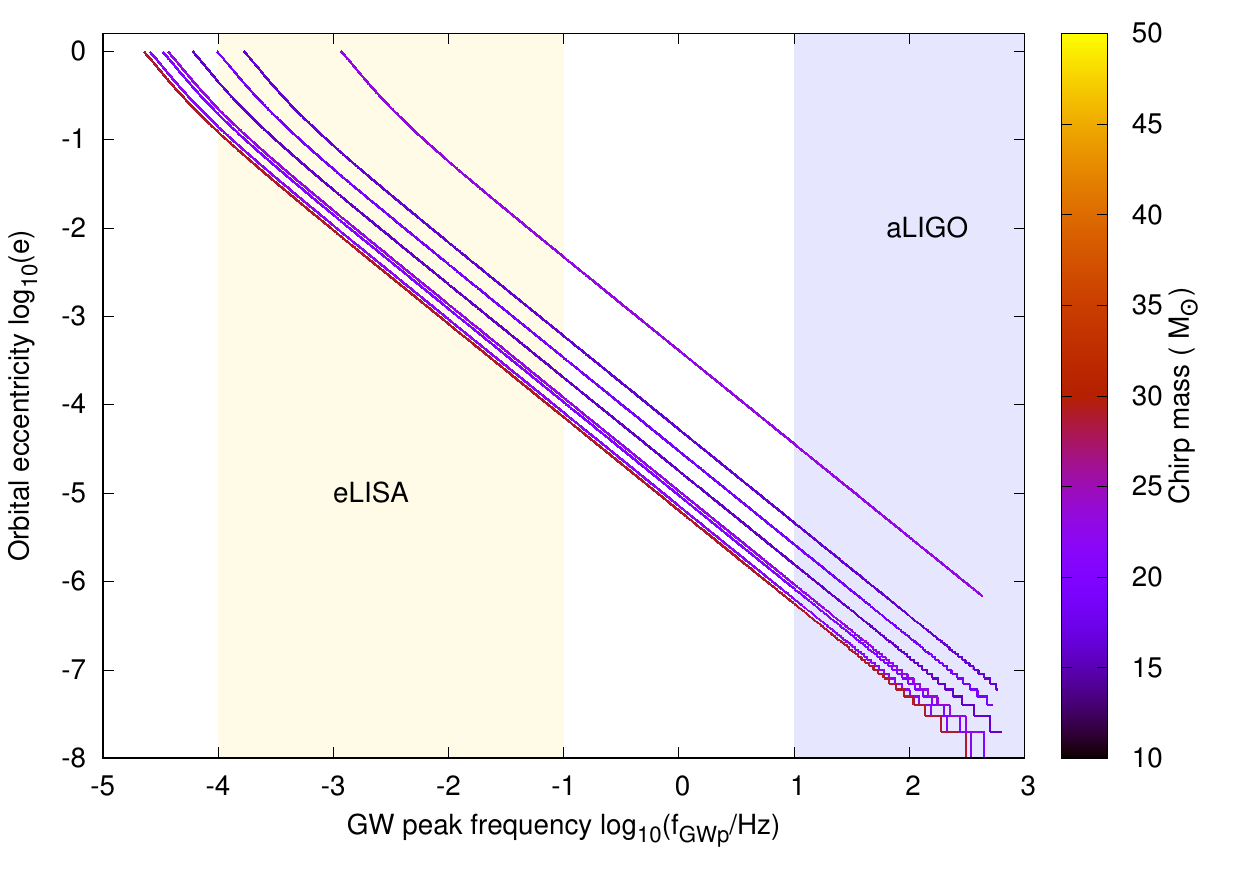}
	\caption{{\bf Top:} Final orbital inspiral curves of the coalescing BBHs, that
	are bound to the clusters, in the $\log_{10}(e)-\log_{10}(\fgwp)$ plane,
	$e$ being the BBH's eccentricity
	and $\fgwp$ being the GW frequency at the peak of the BBH's GW power spectrum
	(see text). The gold and the blue shades represent the characteristic detection
	frequency bands of the LISA and the LIGO instruments. The curves are
	obtained by integrating the \citet{Peters_1964} orbit-averaged semi-major-axis and eccentricity
	decay formulae, initiating with the BBH parameters in the coalescence event
	record. They are colour coded according to the corresponding BBH's chirp mass.
	These curves suggest that although they begin spiralling in
	due to very high eccentricities (driven by resonant triples/Kozai mechanism; see text), the BBHs
	typically become sufficiently circular to be detectable by the LISA (see text) whilst
	traversing the instrument's frequency band and become nearly circularized
	by the time they enter the LIGO's frequency band.
	{\bf Bottom:} The same, for the ejected BBHs that coalesce within a Hubble
	time and a similar description applies to them; they acquire the high
	eccentricities, at the beginning of their in-spiral, due to the close
	dynamical encounters that have led to their ejections.
	}
\label{fig:inspiral}
\end{figure*}

\subsubsection{Binary black hole inspirals in the computed models}\label{inspiral}

Nearly all BBH inspirals in the present models, be it triple-induced or ejected,
begin with very high eccentricities, that ensure their short coalescence times
(Eqn.~\ref{eq:taumrg}). For the triple-induced (bound) cases, the
resonant interaction/Kozai mechanism (Sec.~\ref{intro})
induces a high eccentricity in the inner BBH and, in the ejected BBHs, 
the high eccentricities are due to their internal orbital modifications by the close
encounters that ejected them. While a high $e$ ensures short $\taumrg$ or at
least $\taumrg<$ Hubble time, it also raises the question whether a detectable
eccentricity remains when the in-spiralling BBH emits GW within, say, the LIGO's detection
band or whether such eccentric BBHs will be audible (as persistent sources) by the future space-based
GW observatories like the Laser Interferometer Space Antenna (hereafter LISA; \citealt{eLISA}).

The GW radiated by an eccentric binary is broadband whose peak frequency is given by
\citep{Wen_2003}
\begin{equation}
\fgwp=\frac{\sqrt{G(m_1+m_2)}}{\pi}
	\frac{(1+e)^{1.1954}}{\left[a(1-e^2)\right]^{1.5}}.
\label{eq:gwfreq}
\end{equation}

\begin{figure*}
\centering
\includegraphics[width=8.5cm,angle=0]{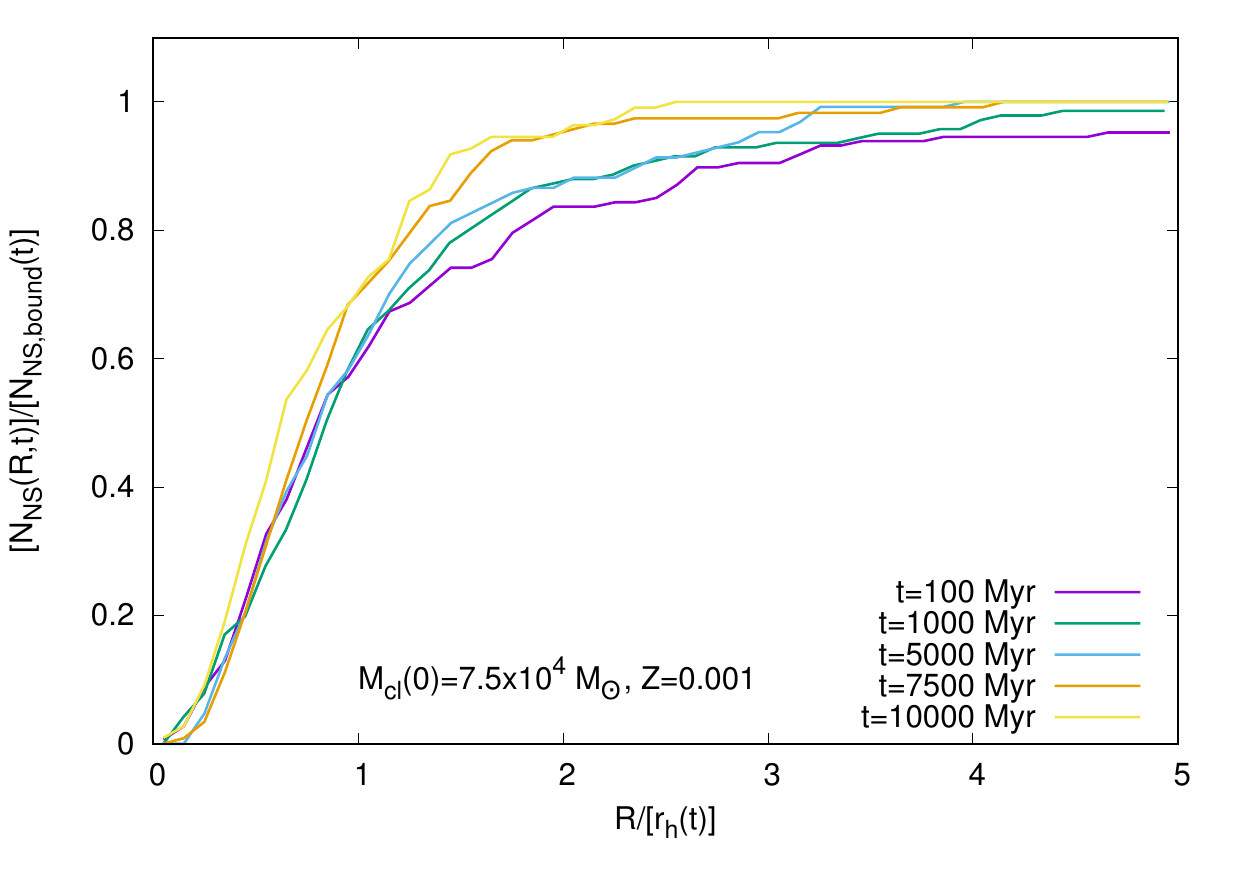}
\includegraphics[width=8.5cm,angle=0]{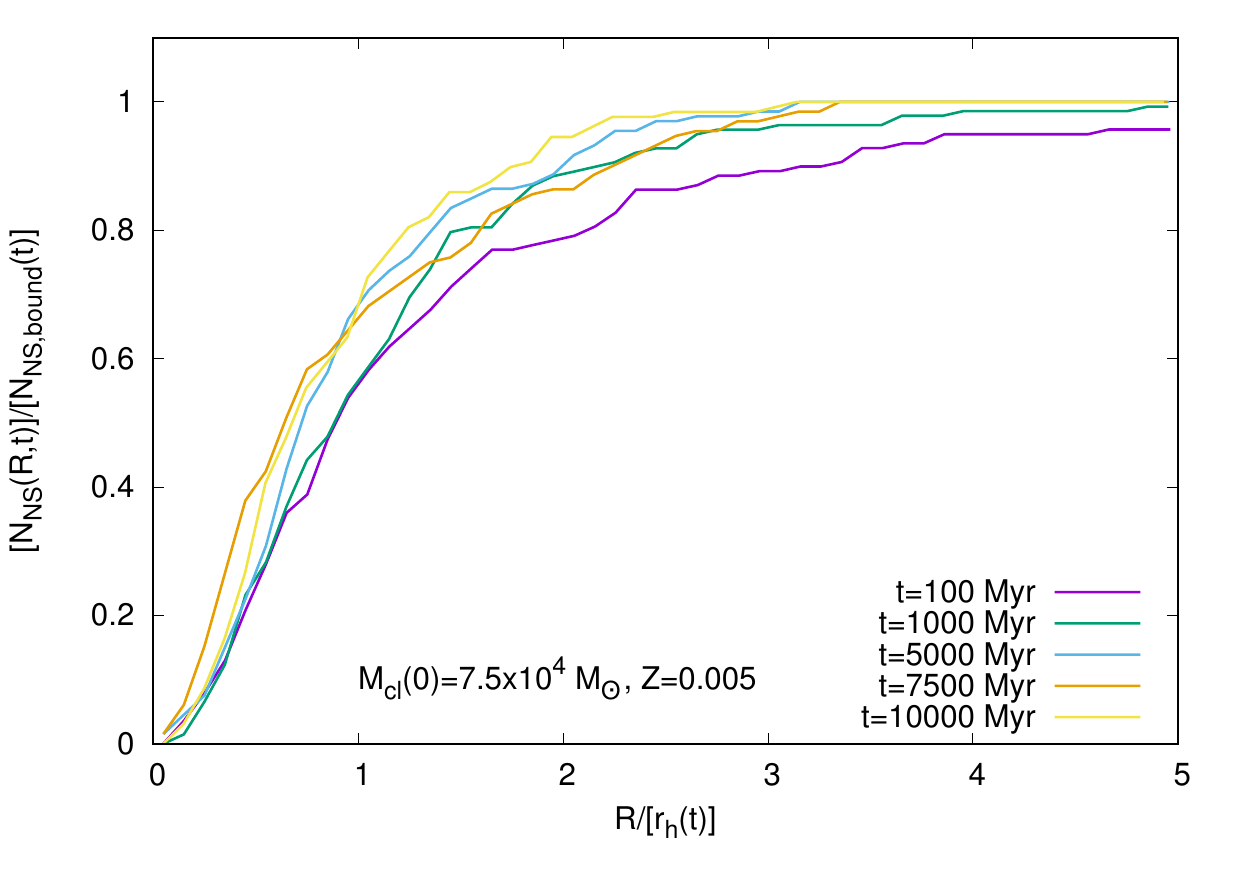}
\includegraphics[width=8.5cm,angle=0]{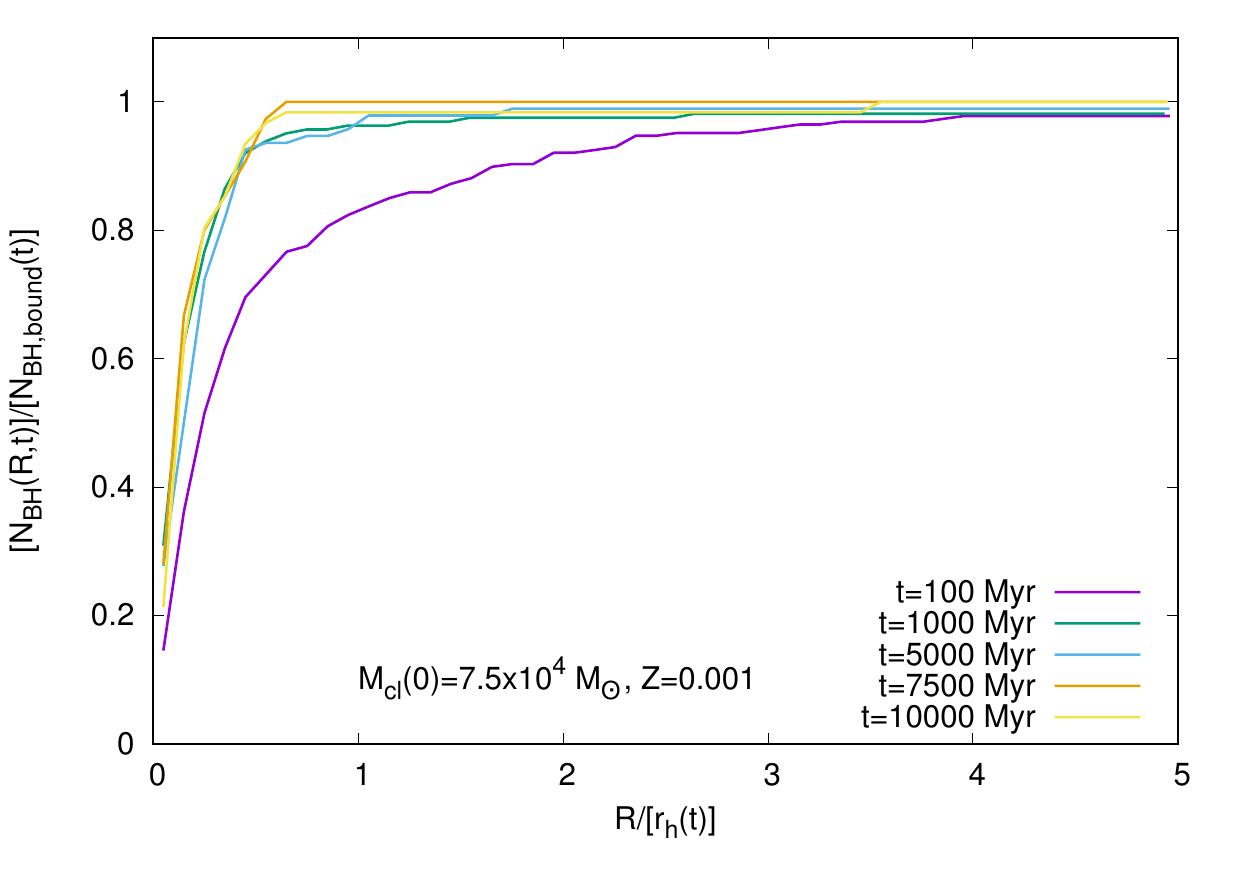}
\includegraphics[width=8.5cm,angle=0]{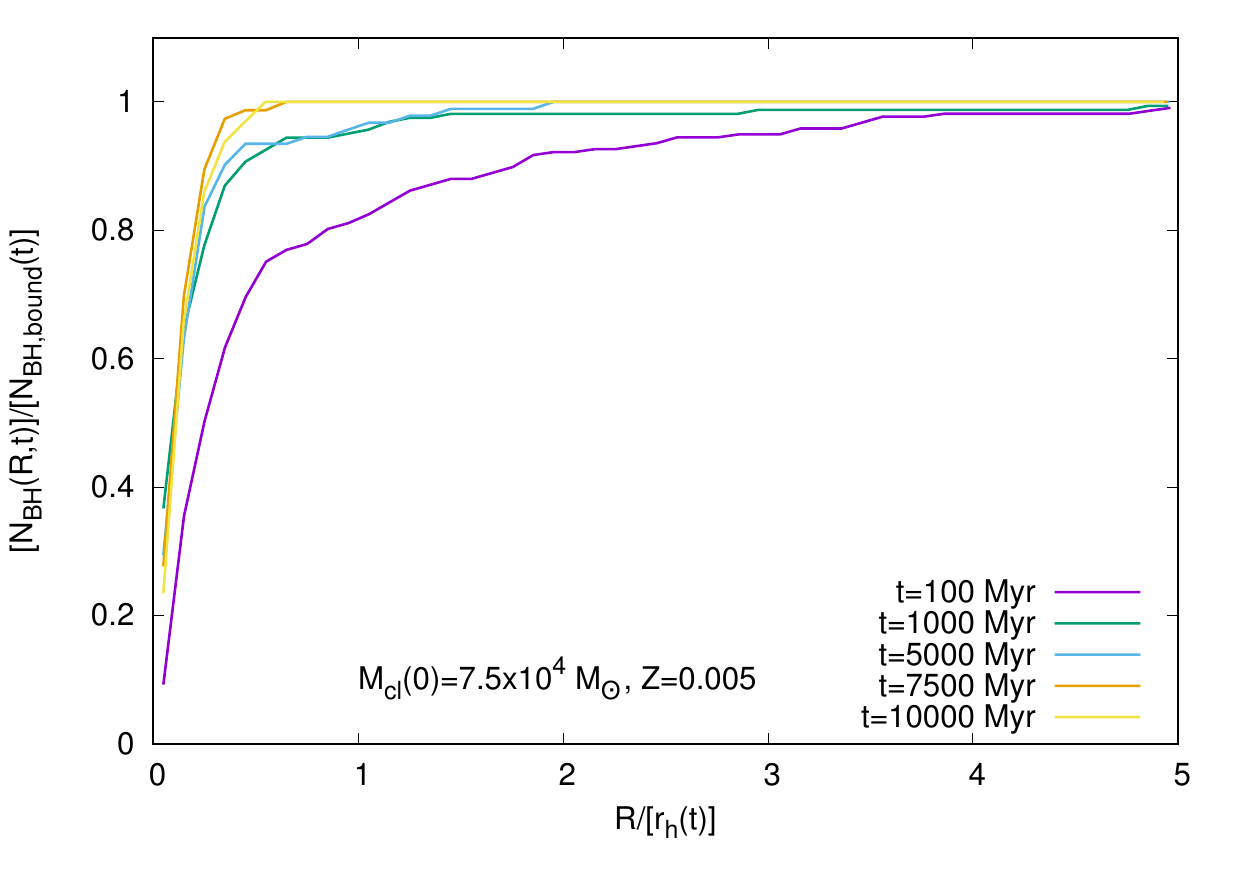}
	\caption{{\bf Top:} The cumulative radial distribution of the (bound) NSs, at increasing
	evolutionary times (legends), for the $\mcl(0)\approx7.5\times10^4\Ms$ computed models
	(Table~\ref{tab1}). The number of NSs, $N_{\rm NS}(R,t)$, within a radial distance, $R$,
	and at an evolutionary time, $t$,
	is normalized w.r.t. the total number of NSs, $\nnsbound(t)$, bound to the cluster at that time
	and the radial axis is normalized w.r.t. the cluster's half-mass radius, $\rh(t)$, at time $t$.
	These cumulative distributions imply that although the NSs continue to centrally segregate with time,
        the mass segregation is inefficient due to the dynamical energy injection by the BHs, as already indicated
	in \citet{Banerjee_2017} --- at late evolutionary times, $\approx90$\% of the NSs occupy 2 half-mass radii.
	{\bf Bottom:} The same treatment as in the top panels but for the BHs bound to the clusters.
	They imply that, except for very early evolutionary times, the much more massive BHs are generally
	much more strongly centrally concentrated than the NSs,
	as can be expected --- nearly all BHs are contained within one
	half-mass radius.} 
\label{fig:nsdist_75k}
\end{figure*}

Fig.~\ref{fig:inspiral} (top panel) shows the trajectories,
in the $\log_{10}(e)-\log_{10}(\fgwp)$ plane,
of the in-spiralling inner BBHs of the BH triples, that have merged while being
bound to the clusters in Table~\ref{tab1} (see its column `f'). At this
stage, the GW radiation timescale of the inner BBH is much shorter compared to the Kozai/dynamical
timescale of the associated triple (as indicated by the ARC treatment;
see Sec.~\ref{nbprog} and references therein), making
the GR coalescence unavoidable. These tracks are obtained by
integrating the \citet{Peters_1964} GR semi-major-axis and eccentricity decay formulae,
beginning with the orbital parameters of the merging BBHs that are logged by $\nbseven$ in
its GR coalescence events' records\footnote{A better way to obtain such trajectories
of triple-induced GR inspirals would be to allow
for an on-the-fly, high-time resolution output of the orbit of a BBH during its GR inspiral,
from within $\nbseven$, which is planned in a future paper. That way, any modifications of the orbital
decay due to the perturbations from the third body could also be tracked.}. For most of these BBHs,
when they have high eccentricities
they emit GW with $\fgwp$ either within the LISA's characteristic detection frequency band
or within the deciHertz range (see Fig.~\ref{fig:inspiral}, top panel). 
Most of the BBHs which initially emit in the LISA band,
circularize sufficiently while traversing through the same band so that
they eventually become audible by the LISA (when $e\lesssim0.7$;
see, \eg, \citealt{Chen_2017}) and their remaining eccentricities
would then serve as signatures of their dynamical origin \citep{Nishi_2016,Nishi_2017}.
By the time the BBHs spiral in to emit
(in terms of $\fgwp$) within the LIGO's
characteristic detection frequency band, they are practically circularized; the most eccentric
BBHs in the LIGO band have $10^{-2}<e<10^{-1}$. For all these BBHs, the time spent
in the plotted trajectories until merger vary from a few years to $\approx0.1$ Myr.    
Hence, some of these BBHs would qualify for the combined LISA-LIGO detectability \citep{Sesana_2016}.

Fig.~\ref{fig:inspiral} (bottom panel) shows such tracks for the escaped BBHs
with $\taumrg<13.7$ Gyr, from all the computed clusters in Table.~\ref{tab1} (see its column `g'). 
As in the panel above, these tracks are obtained by integrating the \citet{Peters_1964}
formulae, initiating with the orbital parameters of these BBHs at the instants
of their ejections (as logged by $\nbseven$). All of these BBHs are ejected
with very high eccentricities (which is why they have shorter $\taumrg$; see Eqn.~
\ref{eq:taumrg}) when their $\fgwp$s lie within or below the LISA's detection
band. As in the case of triple-induced mergers (see above), these BBHs
circularize sufficiently to be heard by the LISA while traversing through the
instrument's detection band and are practically circularized when they enter
the LIGO's detection band.

Studies such as
\citet{Samsing_2014,Samsing_2017,Antonini_2014,Antonini_2016b,Antonini_2017}
suggest the occurrences of BBH mergers
that retain significant eccentricities, $e>0.1$, in the LIGO band ($\fgwp\geq10$ Hz);
matching with the corresponding GW waveforms
would require GW templates from eccentric binaries \citep{Huerta_2013}.
In the latter three studies, the
same ARC code as here (Sec.~\ref{nbprog}) has been applied to
directly integrate the (isolated/field) triple BHs. Note
that the GC- or nucleus-type model clusters, that are considered
as the hosts/parents of the triple-BHs/ejected-BBHs in the above works,
are about an order of magnitude more massive
than the present open cluster-type models. Therefore, the former clusters
would contain tighter BBHs that can become
members of triples or be ejected, with higher probabilities of retaining higher eccentricities
in the LIGO band. With a larger model set giving a larger sample of triple-induced and ejected
BBH mergers, such ``eccentric'' LIGO mergers can, perhaps, be obtained
in similar models.

\subsection{Properties of the neutron star population: mass segregation
and neutron star binaries}\label{nsseg}

\begin{figure*}
\centering
\includegraphics[width=8.5cm,angle=0]{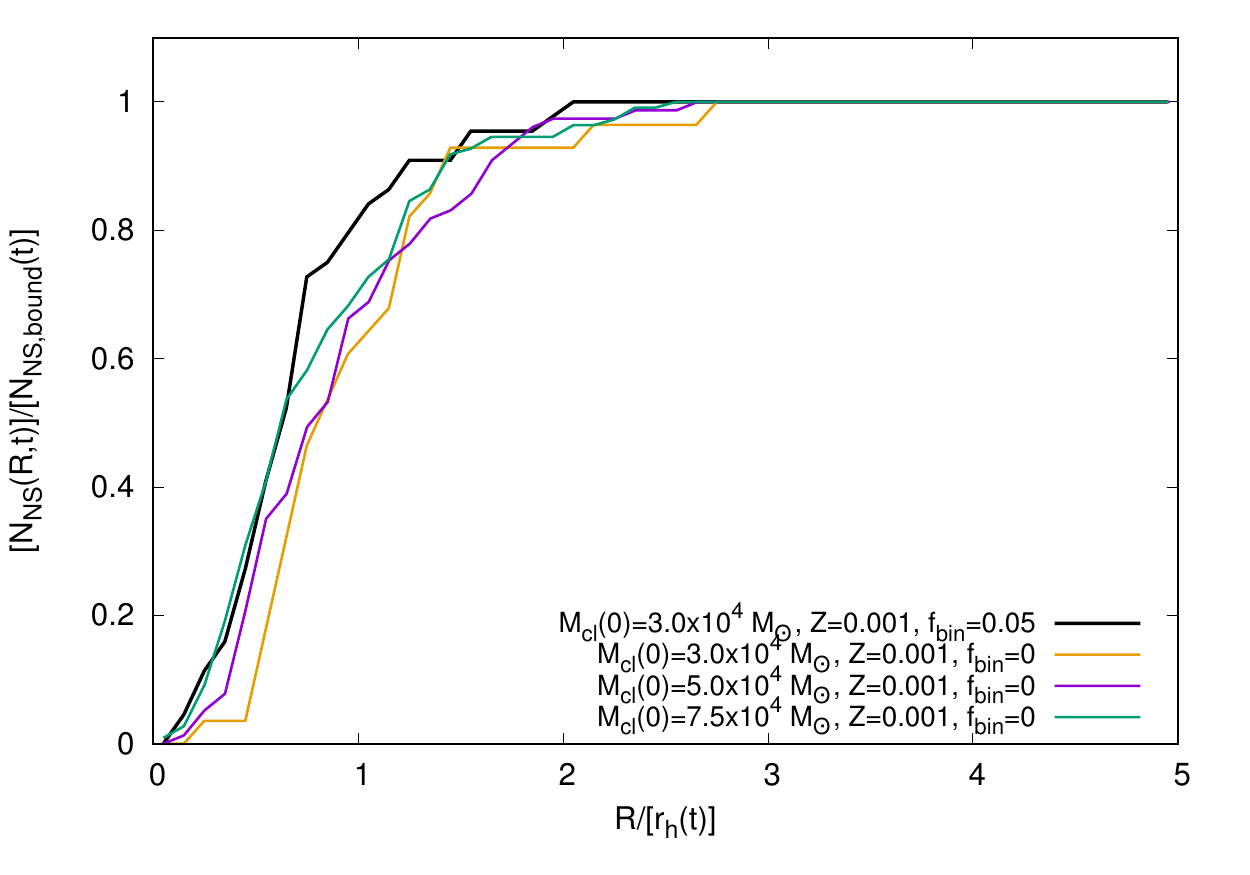}
\includegraphics[width=8.5cm,angle=0]{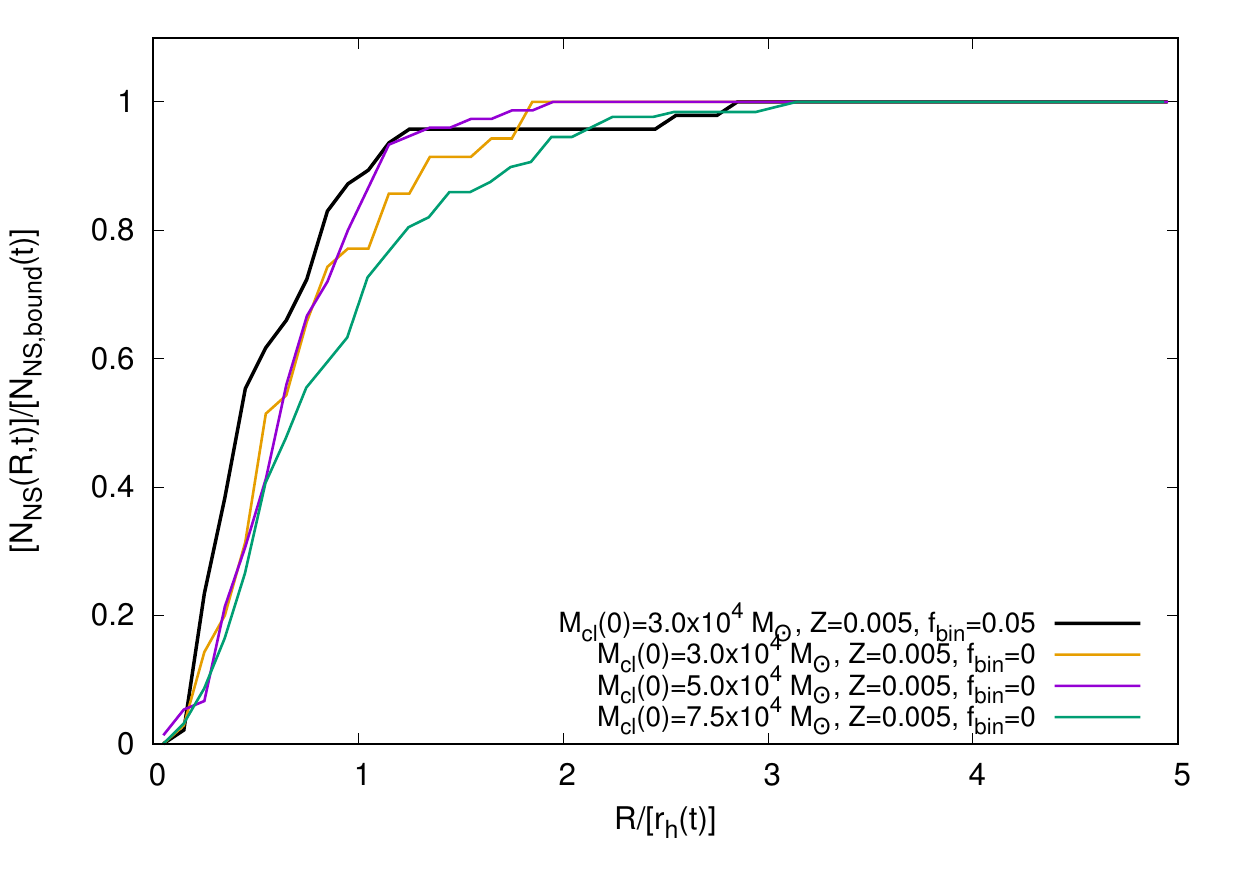}
	\caption{The normalized cumulative radial distribution of NSs, as in Fig.~\ref{fig:nsdist_75k},
	at $\approx10$ Gyr evolutionary time and for the models with
	$3.0\times10^4\Ms\lesssim\mcl(0)\lesssim7.5\times10^4\Ms$ (legends) and
	$Z=0.05\Zs$ (left panel) and $0.25\Zs$ (right panel). Those for the computed models
	with $\fbin\approx5$\% overall binary fraction (Table~\ref{tab1}) are shown
	in the black lines. A higher rate of BH depletion, in the presence
	of primordial binaries (Fig.~\ref{fig:comp_30k}; Sec.~\ref{res}), aids the mass segregation
	of the NSs, making them somewhat more centrally concentrated at 10 Gyr, compared
	to the cases with no primordial binaries.}
\label{fig:nsdist_10Gyr}
\end{figure*}

The sustained energy injection by the BHs modifies the standard two-body relaxation driven
processes in the cluster such as its approach towards core collapse and mass segregation.
In particular, as demonstrated in Paper I, the presence of the BHs would keep the NSs from
undergoing mass segregation significantly and thus from becoming a subpopulation that, like the BHs, would   
control the dynamics of the cluster and form relativistic subsystems
efficiently (see Figs.~6 \& 7 of Paper I and the corresponding discussions therein).
Fig.~\ref{fig:nsdist_75k} (top panels) illustrates this explicitly with the
$\mcl(0)\approx7.5\times10^4\Ms$ models; w.r.t. the cluster's (time-varying)
half-mass radius, $\rh(t)$, the NSs' central concentration increases only moderately
with time. At the latest evolutionary times, the NSs occupy the volume engulfed by
$R=3\rh(t)$ with $\approx90$\% of the NSs lying within $R<2\rh(t)$. In contrast, except
for the earliest evolutionary times when it is still in the midst of mass segregation,  
the bulk of the BHs always remains concentrated well within $\rh(t)$ (Fig.~\ref{fig:nsdist_75k},
bottom panels). Similar descriptions are valid for the long-term evolution of the NSs'
radial distribution for the other lower-mass models in Table~\ref{tab1}, as illustrated
in Fig.~\ref{fig:nsdist_10Gyr}; this figure also demonstrates that the more efficient
ejection of the BHs from the clusters with primordial binaries allows a marginally higher
degree of segregation of the NSs in them (the black lines in Fig.~\ref{fig:nsdist_10Gyr}).

Notably, the majority of the NSs in GCs are found well within the clusters' half-mass
radii (see, \eg, \citealt{rsm2008,rsm2008b,hv83}),
unlike what is derived here. This may not, however, represent a discrepancy   
since such pulsar surveys in GCs are unlikely to reveal the complete NS populations and
could be biased towards the clusters' innermost regions where the NSs' spatial densities  
are still the highest. A direct comparison, in this respect, between the present models
and much more massive GCs would not hold; a much higher stellar density and hence
stronger dynamical friction in GCs would aid the NSs' segregation, resulting in more
centrally-concentrated NS profiles.

The present $\fbin>0$ models in Table~\ref{tab1} are found to produce any
hard double neutron star (hereafter DNS) binary\footnote{In this work, any hard binary
between two NSs or any binary containing an NS will be regarded as an NS binary,
irrespective of the NSs' recycling status. Non-recycled NS binaries, such as those
assembled dynamically or formed via detached evolution of massive primordial binaries,
have limited possibilities to be caught through radio observations.}
\emph{neither} through dynamical means \emph{nor} through the evolution
of massive (dynamically-active; see Sec.~\ref{nbprog}) primordial binaries. Note that although
the formation of DNSs through the former channel depends on the extent of the NSs' dynamical activity,
which, in turn, depends on how well concentrated they are, 
the latter channel always have the potential to give rise to DNSs in stellar clusters
(see, \eg, \citealt{Belczynski_2002,Lorimer_2008,Belczynski_2010b,Oslowski_2011,Tauris_2017}).
The non-occurrence of DNSs, in the present lower-mass ($\mcl(0)\approx3.0\times10^4\Ms$),
primordial-binary models, is a combined consequence of
(i) an overall small population of potential DNS-progenitor (primordial) binaries in such lower-mass models
with a low primordial-binary fraction ($\fbin<0.1$ for $\mzams<16\Ms$; see Sec.~\ref{calc}), 
(ii) dissociation of all detached binaries, where each member would evolve into an NS,
due to the associated large supernova mass loss, 
(iii) the stellar-wind, its tidal modification, and mass-transfer schemes adopted in the
\bse (see Sec.~\ref{nbprog} and references therein)
that determines the pre-supernova (adiabatic) mass loss (\eg, through CE ejection,
regular and tidally-enhanced winds) which is instrumental in the post-supernova survival of
the binary and its members' recycling, (iv) the implementation of the NSs' natal kicks in the \bse,
and (v) dynamical perturbations of
the potential DNS-progenitor binaries. The evolution of massive binaries in a
dynamically-active environment and the consequent formation of compact binaries has not
yet been well explored and is planned in a future study.

Note that even if
a DNS forms through either of the channels, its members are prone to get exchanged with
much more massive BHs, when the latter are ambient, limiting the lifetime
of the DNS in such an environment; see below. In the present models,
the only type of double-compact binaries, that is derived directly from
(dynamically-processed) primordial binaries, is the NS-BH binaries (see below).

\begin{figure*}
\centering
\includegraphics[width=8.5cm,angle=0]{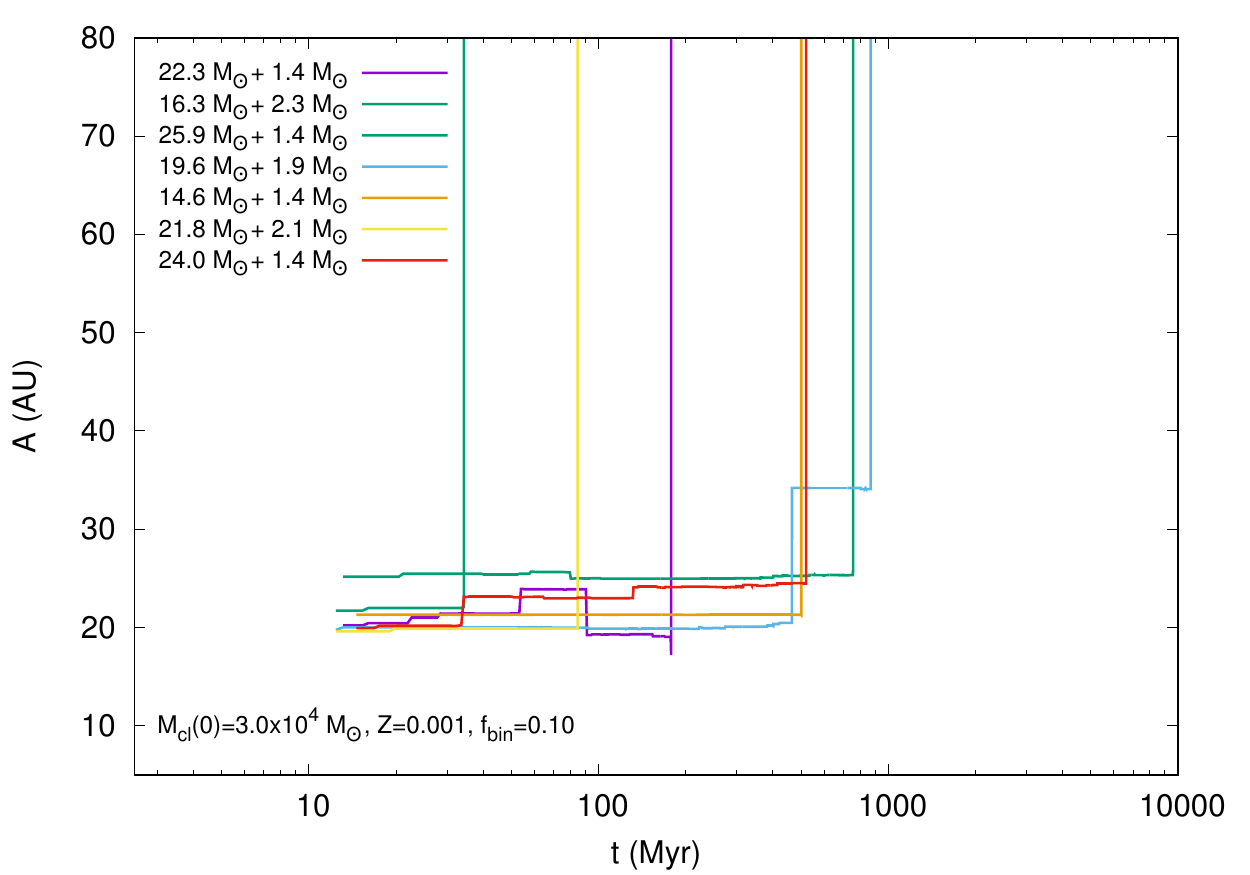}
\includegraphics[width=8.5cm,angle=0]{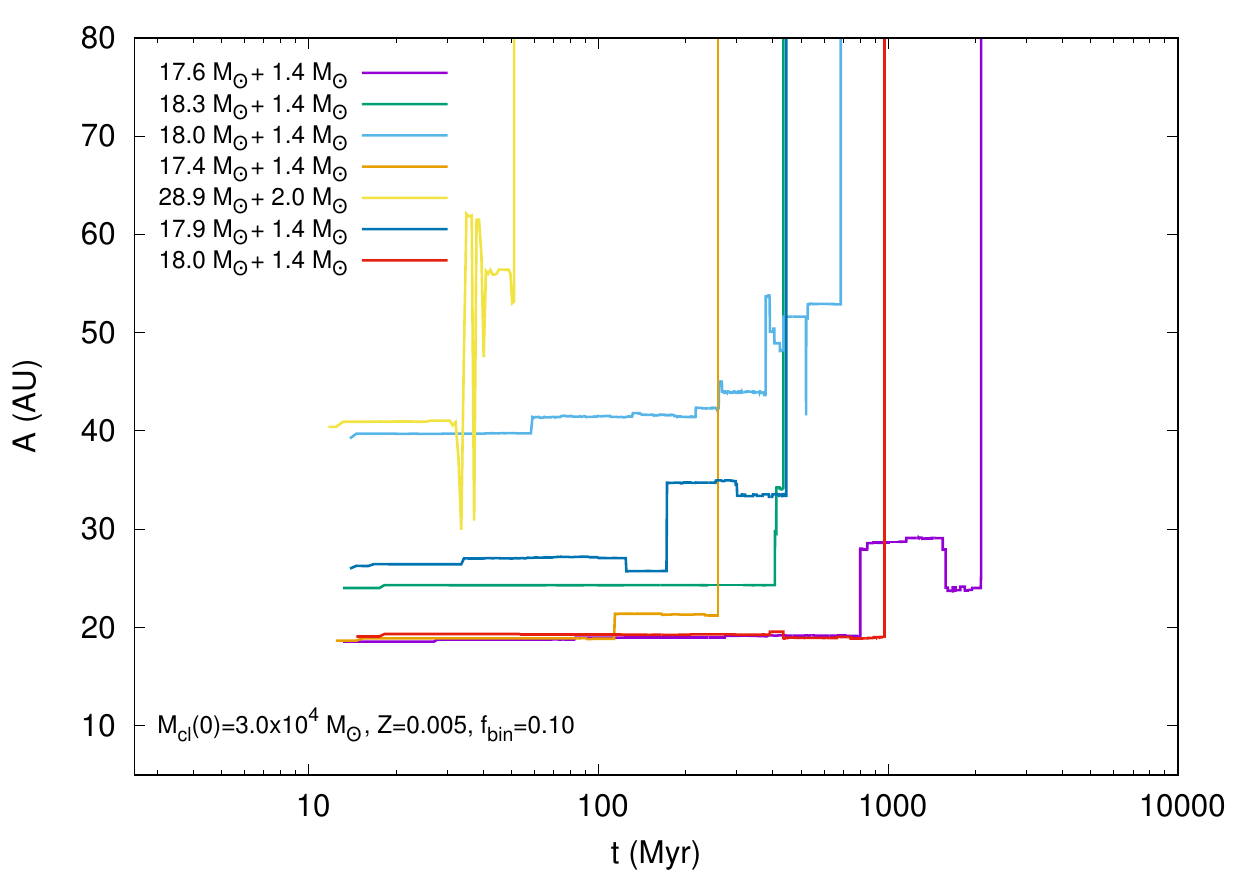}
\caption{The time evolution of the semi-major-axes, $A$, of the NS-BH binaries present in the
	computed models with $\mcl(0)\approx3.0\times10^4\Ms$ and $\fbin\approx10$\%
	(Table~\ref{tab1}). Each line follows a particular NS-BH binary, with
	member masses as indicated in the legends. All NS-BH binaries, that appear
	in these models (as well as in the other ones with primordial binaries; see Table~\ref{tab1}),
	are formed through the evolution of (possibly dynamically-modified) primordial binaires.
	Their lifetimes, in the clusters, vary from $\sim100$ Myr to $\sim$ Gyr,
	by which time the pairs get destroyed, often through exchange interactions (Sec.~\ref{nsseg}).
	Typically, such binaries are born several 10s of AUs wide
	and they retain sizes of that order until their destructions
	(true for the NS-BH binaries in the other primordial-binary models also).}
\label{fig:nsbh_semi}
\end{figure*}

Each of the $\fbin>0$ models are found to contain several NS-BH binaries (and to
rarely eject NS-BH binaries). Nearly all of such binaries are derived directly  
from a primordial massive binary (Sec.~\ref{calc}), \ie, both the NS and the BH
are derived from progenitors that were members of the same primordial pair.  
Fig.~\ref{fig:nsbh_semi} shows the semi-major-axes of such NS-BH binaries,
from the point of their appearance, for the $\fbin\approx10$\% models (Table~\ref{tab1}).
Here, the semi-major-axis, $A$, is simply the relative separation between the NS and the BH
member obtained at each \nbseven output time, representing the width of the binary.
Each of these NS-BHs are formed through detached evolution of a tight primordial binary
that has avoided exchange (as indicated by the serial IDs of the NS and the BH members
in the outputs), \ie, they are non-recycled.
The formation of the (moderately-massive) BH member through direct collapse
(see Secs.~\ref{intro} \& \ref{nbprog} and references therein) has enabled the binary
to survive the supernova mass loss, as opposed to while forming a DNS (the other factors,
as raised above, being common), giving the resulting BH-NS a higher chance of survival.
As for the DNS (non) formation (see above), it is impossible to capture all
possible scenarios of NS-BH formation in these models, given that only a handful
of the potential progenitors would exist per model.

As seen in Fig.~\ref{fig:nsbh_semi}, the widths of the NS-BHs, whilst
they remain bound to the cluster, continue to alter due to the close encounters they experience.
Their widths ultimately blow up, when, typically, the NS member gets exchanged
by an intruding BH and is ejected from the binary. Such an exchange encounter
limits the lifetime of an NS-BH, which, in the present models, lie between $\sim 100$ Myr
and $\sim$ Gyr (Fig.~\ref{fig:nsbh_semi}); in a more massive cluster with a denser
but higher-velocity dispersion BH population, the lifetime is expected to be similar
\footnote{The encounter rate as seen by a binary, $\gamma\propto\rho/\sigma$ \citep{Banerjee_2006}, where
$\rho$ and $\sigma$ are the ambient stellar (BH) density and velocity dispersion,
respectively. Applying the virial theorem, since both $\rho$ and $\sigma\propto M$, the (BH) cluster
mass, $\gamma$ is independent of $M$ and depends only on the cluster's scale length.}.
This should also
apply to any DNS that happens to be present in the cluster, where the NS members are likely get exchanged
sequentially by the BHs, within $\sim$ Gyr time. Likewise the DNS case, no NS-BHs, in
these models, are derived through the dynamical channel except in 1-2 cases
the original BH member is exchanged with an intruder BH. In the $\fbin\approx0.1$,
$Z=0.05\Zs$ model, one of the NS-BH is found to undergo GR coalescence
(with an exchanged BH member; see Fig.~\ref{fig:bbhmrg}) after a sequence of close encounters and
ultimately triple formation with an intruder BH. Such NS-BH mergers would act
as both GW sources for the LIGO and short gamma ray burst (sGRB) sources. 

Note that the presence of any type of compact binary, that is derived through
the internal evolution of a primordial pair (or of an exchanged stellar pair),
in simulations as those here, should be taken with caution as their occurrence
is dependent on the adopted stellar- and binary-evolution schemes --- in the present case,
on the \bse and its integration into \nbseven. Nonetheless, the small
population of NS-BHs, present in the $\fbin>0$ models, could be used as probes
to robustly show that the
NS member of an NS binary is likely to get exchanged by a BH, within $\sim$ Gyr after
the formation of the NS binary (primordially or dynamically), when a dynamically-dominant
BH population is present.

In this work, only hard double-compact binaries have been probed. Despite
the near inhibition of mass segregation of the NSs (and also of other stellar members;
see, \citealt{Alessandrini_2016,Peuten_2016}) due to the work of the BHs, as discussed above, they would,
in principle, undergo exchange encounters with the normal-stellar binaries, potentially forming
progenitors of X-ray binaries. Since the BHs continue to interact with the normal
stars and their binaries (see Secs.~\ref{intro}, \ref{res} \& \ref{bbh}), they have
the potential to acquire a stellar companion as well. Both NS-star and BH-star binaries
and their possible mass-accretion phase thereafter, of course, can also appear through the
internal evolution of the same primordial pair (or of an exchanged stellar pair).
Finally, cataclysmic variables (CV) can appear
within the models via both dynamical and binary-evolution channels. Again, as in the
case of double-compact binaries (see above), a dynamically-dominant population
of BHs would have the potential to quench the exchange activities of
the NSs and the WDs. A more thorough analysis procedure for tracking all types
of compact binaries from a larger set of such simulations is underway.

\subsection{Remark on the dynamical binary black hole merger rate: contribution
from open clusters}\label{mrgrate}

It would be of interest to compare the contribution of BBH merger rate  
from lower-mass, open- or YMC-type clusters, as considered here, with that from classical
GC-type clusters considered in Monte Carlo-based studies (Sec.~\ref{intro}).
Although open clusters and YMCs, in general, produce less BBH mergers per cluster than
that from the GCs, they are more numerous, enabling them to compete with the GCs. 

The present work shows that, like the GCs, the open clusters generate BBH mergers
spanning their entire evolutionary time (Fig.~\ref{fig:bbhmrg}). For definiteness,
the GC models of \citet{Morscher_2015} are considered here which have been utilized by
\citet{Rodriguez_2015} for their BBH merger rate calculations. From Table~\ref{tab1}
here, it can be observed that for the $\fbin=0$ models, increasing 
$\mcl(0)$ by a factor of $\approx8$ increases the average BBH merger count, per cluster, by
$\approx15$; increasing $\fbin$ increases the merger count by a factor of few, for a given
$\mcl(0)$. A similar scaling continues for the GC models in \citet[their Table 3]{Morscher_2015}.
Assuming that clusters are formed with a power-law mass function of index $\approx-2$
\citep{Gieles_2006a,Gieles_2006b,Larsen_2009}, it can be noted that a shift of the lower $\mcl(0)$ limit,
for producing BBH mergers, to $\approx10^4\Ms$ (Table~\ref{tab1}) from the GC-progenitors'
lower mass limit of $2\times10^5\Ms$
\citep{Morscher_2015} would increase the number of potential clusters by $\approx20$.
Based on the above scaling, if the average BBH merger count per cluster is reduced by a
factor of $\approx10$ for the open-cluster/YMC progenitors with $10^4\Ms\lesssim\mcl(0)\lesssim2\times10^5\Ms$,
then, correspondingly, the BBH merger rate from such open clusters will be $\approx2$ times
the GC counterpart. In other words, although YMCs and open clusters
produce less BBH mergers per cluster and live shorter compared to GCs, they, altogether, would add
a comparable amount to the dynamical BBH merger rate.

Note that the above estimate remains valid even after considering the ``delay times'' of
the mergers, since both the open clusters and the GCs continue to produce BHH mergers throughout
their $\gtrsim10$ Gyr lifespan (\cf Fig.~\ref{fig:bbhmrg} here and Fig.~1
of \citealt{2016arXiv160906689C}). In other words, the comparison remains valid even when
the BBH mergers occurring at the current epoch
(within redshift $z\lesssim0.2$) are considered.
Such a semi-quantitative comparison, of course,
depends on the mass threshold of what can be called a progenitor of a GC (here taken
to be $\mcl(0)\approx2\times10^5\Ms$ for specificity). A more proper derivation of the
BBH merger rate, including the open-type clusters, can be done by combining
the BBH mergers from lower-mass direct N-body models as here with those from the massive 
Monte Carlo models, and following the cosmic star (cluster) formation and metallicity evolution
as done in, \eg, \citet{Bulik_2004,Belczynski_2016}. Such a study is planned in the near future
with a more elaborate set of direct N-body models.

\section{Summary}\label{summary}

The study of the dynamics of stellar-remnant BHs in YMCs and open clusters,
that has been initiated in Paper I, is extended in this study to
include more massive clusters ($\mcl(0)\lesssim10^5\Ms$) and primordial binaries (Sec.~\ref{calc};
Table~\ref{tab1}). The key inferences are as follows:

\begin{itemize}

\item For all clusters with initial masses up to $\mcl(0)\lesssim10^5\Ms$ at least and with
	compact sizes (here taken to be $\rh(0)\approx2$ pc), the GR BBH mergers mostly take place
		in situ, within triples that are bound to the cluster
		(Table~\ref{tab1}, Fig.~\ref{fig:bbhmrg}).
	The apparent contrast with the outcomes of Monte Carlo calculations of $\sim10$ times
		more massive clusters, which predominantly yield mergers in BBHs that are
		dynamically ejected from them, is a consequence of the lower-mass open clusters having
		lower escape speeds and hence containing and ejecting generally wider BBHs
		(Sec.~\ref{bbh}).
		However, a difference would also arise due to the more elaborate and complete
		treatments of close encounters and of relativistic, few-body sub-systems, in direct
		N-body approaches as here (Sec.~\ref{nbprog} and references therein).
		In any case, the BBH merger counts per cluster and the delay times
		of the mergers, as obtained here for the open
		clusters, follow qualitatively similar trends with cluster mass, primordial-binary
		content, and BBH parameters as those in the Monte Carlo GC models (Sec.~\ref{mrgrate}
		and references therein). 
		Based on such trends, it can be qualitatively inferred that open clusters
		would contribute to the present-epoch (low-redshift) dynamical BBH merger rate to a similar
		extent as the GCs. 

\item The introduction of a small population of primordial binaries ($\fbin\gtrsim0.05$
		overall binary fraction, with $\fobin\approx1.0$ for the O-stars; Sec.~\ref{calc})   
		in the model clusters tends to increase the number of in situ
		BBH coalescences significantly (Table~\ref{tab1}).
		However, in all the $\fbin>0$ computations here, all the in situ (and also the few
		ejected) BBH mergers involve BHs that were members of different primordial pairs, \ie,
		the coalescing binaries have been assembled, through exchange interactions,
		with independently-born BHs. Hence, if these BHs have finite spins,
                all of these mergers will be spin-orbit misaligned as indicated in the GW170104 event
		(Sec.~\ref{intro} and references therein). Besides spin-orbit misalignment,
                the BBH-merger mass ratio being substantially less than unity, especially
		being $\mbhtwo/\mbhone\lesssim0.7$, would serve as an additional indication
		of a dynamical assembly of the BBH (Sec.~\ref{bbh}, Fig.~\ref{fig:mratio}).

\item Among all the to-date confirmed LIGO GW events (\ie, excluding LVT151012), the limits of GW170104's
	parameters encompass the highest fraction of the BBH mergers obtained here (Fig.~\ref{fig:bbhmrg},
		Sec.~\ref{bbh}).  
		All the coalescing BBHs, obtained here, have high eccentricities (owing to
		resonant triples/Kozai mechanism
		or strong dynamical perturbation while getting ejected)
	when they have been emitting GW in the LISA band or in sub-LISA frequencies.
		They circularize sufficiently while traversing the LISA
	band so that they become potentially audible by the LISA ($e\lesssim0.7$)
		and are nearly circularized in when they reach the LIGO band (Sec.~\ref{inspiral};
		Fig.~\ref{fig:inspiral}). The most eccentric BBHs in the LIGO band, here, have
                $10^{-2}<e<10^{-1}$.

\item The energy injection by the BHs into the host cluster (Secs.~\ref{intro},\ref{res}; see also
		Paper I) inhibits the regular two-body interaction---driven mass segregation
		of less-massive entities; in particular, of the NSs
		(Figs.~\ref{fig:nsdist_75k} \& \ref{fig:nsdist_10Gyr}).
		This prevents the NSs to effectively
		participate in exchange interactions as long as a dynamically-dominant BH
		population is present. Especially, like with the NS-BH binaries
		(non-recycled; see Sec.~\ref{nsseg})
		present in the model clusters here, the NS member(s) in any NS binary would
		get exchanged with an intruder BH in $\lesssim 1$ Gyr
		time (Fig.~\ref{fig:nsbh_semi}, Sec.~\ref{nsseg}). 

\end{itemize}

\section*{Acknowledgements}

SB is thankful to the anonymous referee for helpful comments
that has improved some of the descriptions in this paper.
SB is indebted to Sverre Aarseth of the Institute of Astronomy, Cambridge,
for his efforts in improving {\tt NBODY6/7}, without which this study would not
have been possible.
SB is thankful to Sverre Aarseth also for reading the manuscript
and providing improvement suggestions.
SB is thankful to the computing team of the Argelander-Institut f\"ur Astronomie,
University of Bonn, for their efficient maintenance of the workstations
on which all the computations have been performed.

\onecolumn
\begin{center}
\begin{longtable}{lcclcll}
	\caption[Summary of model calculations]
	{Summary of model evolutionary calculations.
	The columns from
	left to right respectively denote: (a) initial mass, $\mcl(0)$, of the model cluster,
	(b) initial half-mass radius, $\rh(0)$, (c) metallicity, $Z$,
	(d) overall primordial binary fraction, \fbin (see Sec.~\ref{calc}), (e) total evolutionary time, \tevol,
	of the model, (f) the number of 
	(triple-mediated) binary black hole (BBH) coalescences, $\nmrgin$, that occurred within the clusters,
	(g) the number of BBH coalescences (in BBHs that are ejected from the clusters), $\nmrgout$, that occurred
	outside the clusters within the Hubble time. The coalescing masses (in \Ms),
	and the evolutionary times, \tmrg (\tej), of the occurrences of the individual
	coalescences (ejections; in Myr) are given in columns (f) \& (g). This table
	includes the new N-body computations (Sec.~\ref{calc})
	and as well those from \citet[Paper I]{Banerjee_2017}.}\label{tab1}\\

	\hline
	\hline
	\mcl(0)/\Ms     & \rh(0)/pc & $Z/\Zs$ & \fbin & \tevol/Gyr & \nmrgin            & \nmrgout\\
	\hline
	\endfirsthead
        
	\multicolumn{7}{c}%
        {{\bfseries \tablename\ \thetable{} -- continued from previous page}} \\
        \hline
	\hline
	\mcl(0)/\Ms     & \rh(0)/pc & $Z/\Zs$ & \fbin & \tevol/Gyr & \nmrgin            & \nmrgout\\
	\hline
	\endhead

	\hline \multicolumn{7}{r}{{Continued on next page}} \\ \hline
        \endfoot

        \hline \hline
        \endlastfoot

	$1.0\times10^5$ & 2.0       & 0.05 & 0.0  & $1$   & 3 (57.7+28.0; 120.0),   & 0                     \\
	                &           &      &      &       & ~~~(53.1+32.9; 165.1),  &                      \\
	                &           &      &      &       & ~~~(40.3+55.4; 286.3)   &                      \\
	\hline
	$1.0\times10^5$ & 2.0       & 0.25 & 0.0  & $1$   & 4 (37.9+40.5; 102.9),   & 0                     \\
	                &           &      &      &       & ~~~(20.3+37.1; 179.3),  &                      \\
	                &           &      &      &       & ~~~(18.2+32.0; 188.6),  &                      \\
	                &           &      &      &       & ~~~(29.1+34.8; 246.6)   &                      \\
	\hline
	$1.0\times10^5$ & 2.0       & 0.50 & 0.0  & $1$   & 0                       & 1 (24.8+19.0; 625.3) \\
	\hline
	\hline
	$7.5\times10^4$ & 2.0       & 0.05 & 0.0  & $10$  & 7 (40.4+64.2; 56.8),     & 0                     \\
			&           &      &      &       & ~~~(43.0+44.3; 200.4),    &                       \\
			&           &      &      &       & ~~~(24.8+28.6; 543.9),    &                       \\
			&           &      &      &       & ~~~(31.1+22.8; 909.2),    &                       \\
			&           &      &      &       & ~~~(21.3+25.9; 1852.5),    &                       \\
			&           &      &      &       & ~~~(24.4+26.0; 4417.1),    &                       \\
			&           &      &      &       & ~~~(10.6+22.4; 10010.6)    &                       \\
	\hline
	$7.5\times10^4$ & 2.0       & 0.25 & 0.0  & $10$  & 2 (37.8+32.8; 76.5),       & 1 (22.8+27.6; 508.3)   \\
	                &           &      &      &       & ~~~(17.5+10.1; 10298.6)    &                        \\
	\hline
	\hline
	\footnotemark[1]
	$5.0\times10^4$ & 2.0       & 0.05 & 0.0  & $>10$ & 1 (24.3+17.7; 2466.1)     & 1 (26.0+42.8; 603.0)\\
	\hline
	\footnotemark[1]
	$5.0\times10^4$ & 2.0       & 0.25 & 0.0  & $>10$ & 1 (34.5+22.7; 151.5)      & 0                    \\
	\hline
	\footnotemark[1]
	$5.0\times10^4$ & 2.0       & 1.00 & 0.0  & $10$  & 3 (9.0+7.5; 5210.1),      & 0                    \\
			&           &      &      &       &   ~~~(10.6+9.4; 7171.0),  &                      \\
			&           &      &      &       &   ~~~(9.1+9.0; 8117.7)    &                      \\
	\hline
	\hline
	\footnotemark[1]
	$3.0\times10^4$ & 2.0       & 0.05 & 0.0  & $>10$ & 1 (38.1+25.9; 933.8)     & 2 (25.7+13.8; 1828.0), \\
	                &           &      &      &       &                           & ~~~(23.6+22.3; 4464.1) \\
	\hline
	\footnotemark[1]
	$3.0\times10^4$ & 2.0       & 0.25 & 0.0  & $>10$ & 0                         & 2 (35.3+20.3; 1787.1), \\
			&           &      &      &       &                           & ~~~(15.7+12.2; 12463.7) \\
	\hline
	\footnotemark[1]
	$3.0\times10^4$ & 2.0       & 1.00 & 0.0  & $>10$ & 1 (10.6+9.0; 3495.9)   & 0                       \\
	\hline
	\hline
	\footnotemark[1]
	$1.5\times10^4$ & 2.0       & 0.05 & 0.0  & $>10$ & 1 (49.4+30.9; 109.7)    & 0                      \\
	\hline
	\footnotemark[1]
	$1.5\times10^4$ & 1.0       & 0.25 & 0.0  & $5$   & 0                         & 0                      \\
	\hline
	\hline
	\footnotemark[1]
	$1.0\times10^4$ & 2.0       & 0.05 & 0.0  & $9$   & 0                         & 0                      \\
	\hline
	\footnotemark[1]
	$1.0\times10^4$ & 1.0       & 0.05 & 0.0  & $5$   & 1 (43.6+34.5; 153.8)     & 0                      \\
	\hline
	\footnotemark[1]
	$1.0\times10^4$ & 1.0       & 0.25 & 0.0  & $5$   & 0                         & 0                      \\
	\hline
	\hline
	\footnotemark[1]
	$0.7\times10^4$ & 1.0       & 0.05 & 0.0  & $7$   & 0                         & 0                      \\
	\hline
	\hline
	$3.0\times10^4$ & 2.0       & 0.05 & 0.02 & $10$  &  0                        &  1 (25.9+27.2; 1410.9) \\
	\hline
	$3.0\times10^4$ & 2.0       & 0.25 & 0.02 & $10$  &  2 (19.0+17.2; 3513.1),  &  0                    \\
	                &           &      &      &       & ~~~(18.9+16.2; 5185.2)   &                      \\
	\hline
	\hline
	$3.0\times10^4$ & 2.0       & 0.05 & 0.05 & $10$  &  1 (22.4+14.8; 6838.7)    &   0                     \\
	\hline
	$3.0\times10^4$ & 2.0       & 0.25 & 0.05 & $10$  &  4 (33.3+34.8; 490.3),    &  0                    \\
	                &           &      &      &       & ~~~(34.5+33.8; 799.8),  &                      \\
	                &           &      &      &       & ~~~(17.5+22.6; 3368.0),  &                      \\
	                &           &      &      &       & ~~~(17.2+13.5; 6649.7)  &                      \\
	\hline
	\hline
	$3.0\times10^4$ & 2.0       & 0.05 & 0.1  & $10$  &  3 (37.9+36.3; 340.2),   &  0                  \\
	                &           &      &      &       & ~~~(28.0+1.85; 1561.0),  &                      \\
	                &           &      &      &       & ~~~(21.3+13.3; 10025.2)  &                      \\
	\hline
	$3.0\times10^4$ & 2.0       & 0.25 & 0.1  & $6$   &  3 (38.8+12.3; 307.0),  &    0                \\
	                &           &      &      &       & ~~~(18.2+21.2; 383.2),  &                      \\
	                &           &      &      &       & ~~~(17.7+17.6; 5310.3)  &                      \\

\end{longtable}
\footnotetext[1]{From \citet{Banerjee_2017}.}
\end{center}
\twocolumn

%\nocite{*}

\bibliographystyle{mnras}
\bibliography{bibliography/biblio.bib}

\label{lastpage}
\end{document}